\documentclass[12pt]{iopart}

\usepackage{iopams}

\expandafter\let\csname equation*\endcsname\relax

\expandafter\let\csname endequation*\endcsname\relax

\usepackage{amsmath}

\usepackage[dvipsnames]{xcolor}
\usepackage{amsfonts}

\usepackage{amssymb}
\usepackage[ruled, vlined]{algorithm2e}
\usepackage{multirow}

\usepackage{graphicx}

\usepackage[colorlinks=true]{hyperref}
\usepackage[noabbrev]{cleveref}

\usepackage{harvard}
\usepackage{caption}
\usepackage{subcaption}
\usepackage{placeins}
\usepackage[normalem]{ulem}
\usepackage{color}

\newcommand\eg{e.g.\ }
\newcommand\ie{i.e.\ }

\begin{document}
\hypersetup{
    urlcolor=black,
    citecolor=blue,
    linkcolor=black,
    }

\title{Physics-informed deep-learning applications to experimental fluid mechanics}

\author{Hamidreza Eivazi$^{\rm a}$, Yuning Wang$^{\rm b}$ \& Ricardo Vinuesa$^{\rm b, c, *}$}

\address{$^{\rm a}$Institute for Software and Systems Engineering, Clausthal University of Technology, 38678 Clausthal-Zellerfeld, Germany\\}
\address{$^{\rm b}$FLOW, Engineering Mechanics, KTH Royal Institute of Technology, SE-100 44 Stockholm, Sweden\\}
\address{$^{\rm c}$Swedish e-Science Research Centre (SeRC), Stockholm, Sweden}

\ead{he76@tu-clausthal.de, rvinuesa@mech.kth.se}
\vspace{10pt}

\begin{abstract}

    High-resolution reconstruction of flow-field data from low-resolution and noisy measurements is of interest due to the prevalence of such problems in experimental fluid mechanics, where the measurement data are in general sparse, incomplete and noisy. Deep-learning approaches have been shown suitable for such super-resolution tasks. However, a high number of high-resolution examples is needed, which may not be available for many cases. Moreover, the obtained predictions may lack in complying with the physical principles, \eg mass and momentum conservation. Physics-informed deep learning provides frameworks for integrating data and physical laws for learning. In this study, we apply physics-informed neural networks (PINNs) for super-resolution of flow-field data both in time and space from a limited set of noisy measurements without having any high-resolution reference data. Our objective is to obtain a continuous solution of the problem, providing a physically-consistent prediction at any point in the solution domain. We demonstrate the applicability of PINNs for the super-resolution of flow-field data in time and space through three canonical cases: Burgers’ equation, two-dimensional vortex shedding behind a circular cylinder and the minimal turbulent channel flow. The robustness of the models is also investigated by adding synthetic Gaussian noise. Furthermore, we show the capabilities of PINNs to improve the resolution and reduce the noise in a real experimental dataset consisting of hot-wire-anemometry measurements. Our results show the adequate capabilities of PINNs in the context of data augmentation for experiments in fluid mechanics.
    
\end{abstract}

%
%
%
\maketitle
%
%
\section{Introduction}
\label{sec:Introduction}

Deep learning (DL) has brought new opportunities for modeling and forecasting the dynamics of complex multiscale systems due to its excellent performance in non-linear function approximation from examples. The emergence of various architectures,~\eg convolutional neural networks (CNNs)~\cite{Lecun1998} and long short-term memory (LSTM)~\cite{hochreiterschmidhuber}, has led to the development of a variety of DL-based frameworks for modeling complex physical systems by accounting for spatio-temporal dependencies in the predictions. Deep learning has been employed in many scientific and engineering disciplines, \eg astronomy \cite{Waldmann2019}, climate modeling \cite{ham2019}, solid mechanics \cite{Mianroodi2021}, chemistry \cite{Segler2018} and sustainability~\cite{VinuesaNat,Larosa}. Fluid mechanics has been one of the active research topics for the development of innovative deep-learning-based approaches \cite{kutz2017,duraisamyetal,bruntonetal2020,vinuesa2021}. Successful application of deep neural networks (DNNs) has been shown in the data-driven turbulence closure modeling \cite{ling2016,mlrans}, prediction of temporal dynamics of low-order models of turbulence \cite{srinivasan,EIVAZI2021}, extraction of turbulence theory for two-dimensional decaying isotropic turbulence \cite{jimenez2018}, non-intrusive sensing in turbulent flows \cite{guastoni2020,guemes2021}, and active flow control through deep reinforcement  learning~\cite{rabault_2019,RL2020,fluidsRL,guastoni_drl}.

The first example of learning in experimental fluid mechanics dates back to the early 1960s for drag reduction \cite{rechenberg1964}. In the early 1990s several approaches were developed based on neural networks with the application in trajectory analysis and classification for particle tracking velocimetry (PTV) \cite{Adamczyk1988} and particle image velocimetry (PIV)\cite{Adrian1984,TeoNNPIV91,grant1995}. In recent decades, image processing and computer vision have experienced a breakthrough with the introduction of deep learning methods, fostering the application of deep learning in processing and enhancing PIV measurements. The majority of the literature has been devoted to improving spatial resolution, recovering missing data, reconstruction of three-dimensional fields from sparse measurements, and post-processing. 

Reconstruction of high-resolution images from low-resolution ones, the so-called super-resolution reconstruction, has been an active research topic in the field of computer science \cite{Wang2021}. The success of deep learning in super-resolution tasks within computer science motivated its application to fluid mechanics and turbulence using CNNs \cite{fukami2019,Liu2020}, super-resolution generative adversarial networks (SRGANs) \cite{guemes2021}, and unsupervised learning based on cycle-consistent GANs (CycleGANs) \cite{kim2021}. Deep learning has been also applied for super-resolution of four-dimensional (4D) flow magnetic-resonance imaging (MRI) data \cite{Ferdian2020}. However, supervised deep-learning-based super-resolution approaches are limited by the need for high-resolution labels for training, \ie a pair set of low and high-resolution fields are needed, which is difficult to obtain from experiments. Recently, \citeasnoun{guemes2021PIV} showed the capability of GANs to generate high-resolution fields from PIV data without having \textit{a priori} any high-resolution field.

Flow data estimation and reconstruction from incomplete data are other common problems in experimental fluid mechanics. Gappy proper orthogonal decomposition (Gappy POD) \cite{Everson95} has been shown suitable for flow-field reconstruction from incomplete data \cite{Bui-Thanh2004}; however, it can not be directly applied to incomplete original data because it requires the spatial coefficients as a basis from the reference data~\cite{bui2003proper_gappy,willcox2006unsteady_gappy}. Recently, \citeasnoun{Morimoto2021} proposed an autoencoder-type CNN for PIV data estimation considering missing regions that can overcome such limitations. Application of machine learning and deep learning has been also shown for flow-field reconstruction from sparse turbomachinery data \cite{Akbari2021}.
Another research direction for augmenting sparse data is to perform three-dimensional fluid data reconstruction from sectional flow fields \cite{Chandramouli2019}.~\citeasnoun{stulov2021neural} proposed a CNN-based architecture including two-dimensional and three-dimensional convolutions to reconstruct three-dimensional data from a few numbers of two-dimensional sections in a computationally friendly manner.
Moreover, the application of deep learning for performing end-to-end PIV has been shown in the experimental-fluid-mechanics community by \citeasnoun{Rabault2017} and motivated by the progress in computer vision regarding the motion-estimation problem \cite{Hui2018,Yang2019,Cai2020,stulov2021neural}. 

Deep-learning methods are powerful modeling tools for data-rich domains such as vision, language, and speech. However, extracting interpretable information and learning generalizable knowledge is still a challenge, especially for domains with limited data available \cite{Rudin2019,Interp}.~Purely data-driven techniques require large datasets for training that may not be available for many engineering problems. Moreover, these models may fit the observational data very well but lack in complying with the fundamental physical laws. Therefore, there is a pressing need for integrating governing physical laws and domain knowledge into the model training, which can act as an informative prior to informing the model about the empirical, physical, or mathematical understanding of the system besides the observational data. Physics-informed neural networks (PINNs), which were introduced by \citeasnoun{Raissi2019}, provide a framework to integrate data and physical laws in the form of governing partial differential equations (PDEs) for learning. This can lead to models that are robust in the presence of imperfect data,~\eg missing or noisy values, and can provide accurate and physically-consistent predictions~\cite{Karniadakis2021}. 

PINNs are well suited for ill-posed and inverse problems related to several different types of PDEs. PINNs were applied to simulate vortex-induced vibrations \cite{raissiviv} and to tackle ill-posed inverse fluid-mechanics problems \cite{raissisci}.~\citeasnoun{NSFnets} showed the applicability of PINNs for the simulation of turbulence directly, where good agreement was obtained between the direct numerical simulation (DNS) results and the PINNs simulation results.~\citeasnoun{eivazi_pinns_pof} employed PINNs for solving Reynolds-averaged Navier--Stokes (RANS) equations without any specific model or assumption for turbulence and using data as the turbulence closure. PINNs were employed also for super-resolution and denoising of cardiovascular-flow MRI data \cite{FATHI2020,Gao2021} and the prediction of near-wall blood flow from sparse data \cite{arzani2021}. Recently,~\citeasnoun{caikarniadakis2021}
proposed a method based on PINNs to infer the full continuous three-dimensional velocity and pressure fields from snapshots of three-dimensional temperature fields obtained by tomographic background-oriented Schlieren (Tomo-BOS) imaging. A detailed discussion on the prevailing trends in embedding physics into machine-learning algorithms and diverse applications of PINNs can be found in the work by \citeasnoun{Karniadakis2021}, and a review of PINNs applications to fluid mechanics was provided by~\citeasnoun{cai_review}.

In this study, we employ PINNs for the super-resolution of flow fields both in time and space from a limited set of measurements without having any high-resolution targets. This is a very important problem that can enhance the accuracy of experimental data in fluid-mechanics studies. Our objective is to learn a functional mapping parameterized by a neural network from spatial $(x)$ and temporal $(t)$ coordinates to the flow parameters, \ie velocity components and pressure, which provides a continuous solution to the problem. We first show the applicability of PINNs for super-resolution tasks in time by solving an ill-posed problem regarding the canonical case of Burgers’ equation. Next, we examine the performance of PINNs in the super-resolution of flow-field data in the presence of noise for the two-dimensional vortex shedding behind a circular cylinder at a Reynolds number of $Re_D= U_{\infty} D / \nu = 100$ based on the cylinder diameter $D$ and the free-stream velocity $U_{\infty}$ where $\nu$ is the kinematic viscosity. Then, we employ PINNs for the super-resolution of noisy data with different resolutions in time and space for a turbulent minimal channel flow at a Reynolds number based on the channel half-height $h$ and the laminar centerline-velocity $U_{\mathrm{cl}}$ of $Re_{\mathrm{cl}} = U_{\mathrm{cl}} h / \nu  = 5,000$. Finally, we show the feasibility of using PINNs for super-resolution and de-noising of a real experimental dataset consisting of hot-wire-anemometry measurements.

One important novelty of this study is the fact that it does not require high-resolution targets to improve the resolution of the measurements, as opposed to computer-vision approaches such as GANs~\cite{guemes2021}. Furthermore, computer-vision-based methods may not necessarily fulfill the underlying governing equations, whereas, in the present framework, the governing equations are part of the loss function. This ensures the compliance of the high-resolution data with the physics of the problem. Another important novel aspect of this work is the treatment of unsteady problems in PINNs. While recent work has documented successful reconstruction of the mean quantities based on unsteady data~\cite{rigas}, in this work we perform the reconstruction of the instantaneous flow fields, based on low-resolution and noisy data. Note that this is done in a turbulent channel, which is a very challenging flow case. Finally, while there is work in the literature discussing de-noising strategies in fluid mechanics using data-driven methods, such as proper-orthogonal decomposition (POD)~\cite{miguel}, the advantage of the current approach is the fact that it requires a lower amount of snapshots, and the data in those snapshots can be incomplete in space. Therefore, the present framework can significantly help to improve experimental measurements in fluid mechanics.

This article is organized as follows: in $\S$\ref{sec:Methodology} we provide an overview of the physics-informed neural networks and discuss the theoretical background; in $\S$\ref{sec:SR} we discuss the application of PINNs for super-resolution of flow-field data in both time and space without having any high-resolution targets; in \S\ref{sec:exp} we discuss the application of the proposed methods to an actual experimental dataset; and finally, in $\S$\ref{sec:conclusions} we provide a summary and the conclusions of the study.

\section{Methodology}
\label{sec:Methodology}

\subsection{Deep neural networks (DNNs)}

Deep neural networks are universal approximators of continuous functions, obtained by composing simple but non-linear transformations at each level (the so-called layer). By the composition of enough such non-linear transformations and having a large training dataset available, a deep neural network can learn very complex functions using the backpropagation algorithm~\cite{LeCun2015}. Multilayer perceptrons (MLPs)~\cite{mlps} are the most basic type of artificial neural network comprised of two or more fully-connected layers of neurons, where each node is connected to all nodes in the preceding and succeeding layers. Let us consider $X$ and $\Psi$ as the input and output spaces and each pair of vectors $(\pmb{\chi},\pmb{\psi})$ as a training example or sample. The neural network aims to find the mapping ${f: X \rightarrow \Psi}$ using the training set such that a loss function $L(f(\pmb{\chi}); \pmb{\psi})$ is minimized. An MLP approximates the function $f$ using the composition of simple transfer functions, \ie linear matrix operations followed by an element-wise nonlinear function called activation function, \eg sigmoidal or hyperbolic tangent. For the layer $i$, the weight matrix $\mathbf{W}_i$ and the bias $\mathbf{b}_i$ perform the linear matrix transformation from layer $i-1$ to the next; then the operation of the activation function $g_i$ leads to the output of the layer as $\pmb{z}_{i} = g_i(\mathbf{W}_i \pmb{z}_{i-1} + \mathbf{b}_i)$. The goal of training an MLP is to determine $\mathbf{W}_i$ and $\mathbf{b}_i$ using training data for the given activation functions.

\subsection{Physics-informed neural networks (PINNs)}
\label{sec:PINNs}

\subsubsection{Model architecture:}
Physics-informed neural networks (PINNs) are deep-learning-based frameworks for solution or discovery of partial differential equations \cite{Raissi2019}. The goal of PINNs is to include the information from the governing physical laws into the training process of a deep neural network. A PINN comprises two parts: an MLP and a so-called residual network, which calculates the residual of the governing equations. In a general set-up dealing with spatio-temporal partial differential equations, the temporal and the spatial coordinates $(t, \mathbf{x})$ are the inputs of the MLP and the solution vector of the PDE system $\mathbf{u} = f(t, \mathbf{x})$ is the output where the function $f$ is parameterized by the MLP. Automatic differentiation (AD) \cite{baydin2018} is utilized to differentiate the outputs $\mathbf{u}$ with respect to the inputs ($t$, $\mathbf{x}$) and formulate the governing equations.~AD can be implemented directly from the deep-learning framework since it is used to compute the gradients and update the network parameters, \ie weights $\mathbf{W}$ and biases $\mathbf{b}$, during the training. We consider partial differential equations of the general form:
\begin{equation}
    \mathbf{u}_t + \mathcal{N}[\mathbf{u}] = 0,~~~\mathbf{x} \in \Omega, t \in [0, T],
    \label{eq:generalPDE}
\end{equation}
where $\mathbf{u}(t, \mathbf{x})$ denotes the solution vector of the PDE system, $\mathbf{u}_t$ is its derivative with respect to time $t$ in the period $[0, T]$, $\mathcal{N}[\cdot]$ is a nonlinear differential operator and $\mathbf{x}$ is the vector of spatial coordinates defined over the domain $\Omega$. Let us consider the residual of the partial differential equations $e(t, \mathbf{x})$ as the left-hand-side of \cref{eq:generalPDE}:
\begin{equation}
    e := \mathbf{u}_t + \mathcal{N}[\mathbf{u}],
    \label{eq:f}
\end{equation}
while the MLP approximates $\mathbf{u} = f(t, \mathbf{x})$. The function $f$ is a composition of simple differentiable operations, which maps the coordinates to the solution, \ie ${f: (t, \mathbf{x}) \rightarrow \mathbf{u}}$. Therefore, it is possible to obtain derivatives of the outputs with respect to the inputs simply using the chain rule on $f$: $\partial \mathbf{u}/ \partial \mathbf{x} = \mathcal{G}(\mathbf{u}, \mathbf{x})$ and $\partial \mathbf{u}/ \partial t = \mathcal{G}(\mathbf{u}, t)$ where $\mathcal{G}$ indicates the chain-rule operation. Therefore, $\mathbf{u}_t$ and $\mathcal{N}[\mathbf{u}]$ in \cref{eq:generalPDE} can be obtained by applying the chain rule for differentiating compositions of functions in the MLP using AD, thus yielding the residual of the governing equations. The aforementioned process for computation of the residuals through the implementation of AD is known as the residual network.~We use the open-source deep-learning framework TensorFlow \cite{tensorflow} to develop our PINN models. TensorFlow provides the ``\texttt{tf.GradientTape}" application programming interface (API) for AD by computing the gradient of a computation with respect to inputs. TensorFlow ``records" the computational operations executed inside the context of a \texttt{tf.GradientTape} onto a so-called ``tape", and then uses that tape to compute the gradients of a recorded computation using reverse-mode differentiation.

\subsubsection{Loss function:} 
\begin{figure}[ht]
    \centering
    \includegraphics[width=0.75\textwidth]{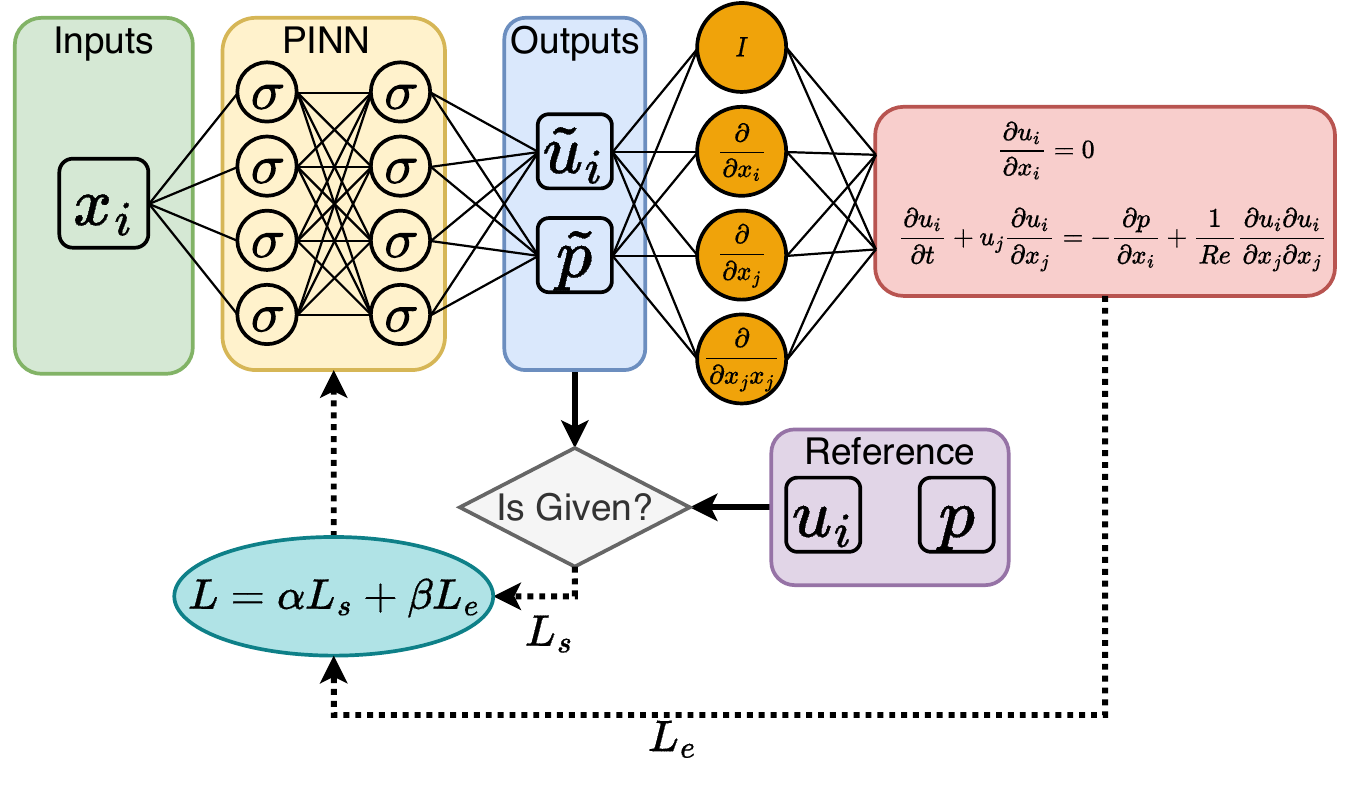}
    \caption{Schematic representation of proposed PINNs architecture. Note the use of the index notation.}
    \label{fig:schematic}
\end{figure}

The training process of a PINN consists of both supervised and unsupervised parts. We refer to supervised learning for those samples for which the target solution is available and unsupervised learning for the samples for which the solution is unknown, while the residual of the governing equations is considered as the loss. Let us consider $N_s$ as the number of points for which we have the solution, \eg initial and boundary conditions or sparse measurements, and $N_e$ as the number of collocation points for which we calculate the unsupervised loss for training. The unsupervised loss can be defined as:
\begin{equation}
    L_e = \dfrac{1}{N_e}\sum_{i = 1}^{N_e}\left| e(t^i_e, \mathbf{x}^i_e) \right|^2,
\end{equation}
where $e(t^i_e, \mathbf{x}^i_e)$ is the residual of the governing equation at each point $\mathbf{x}^i_e$ and instant $t^i_e$, and the supervised loss as:
\begin{equation}
    L_s = \dfrac{1}{N_s}\sum_{i = 1}^{N_s}\left| \mathbf{{u}}_s^i - \mathbf{u}(t_s^i, \mathbf{x}_s^i) \right |^2,
\end{equation}
where the subscript $s$ denotes known values. Furthermore, we define the total loss as:
\begin{equation} \label{eq:loss}
    L = \alpha L_s + \beta L_e,
\end{equation}
where $\alpha$ and $\beta$ are the weighting coefficients.~The shared parameters between the neural network and the residual network can be learned in a physics-informed manner by minimizing the total loss $L$. As discussed by \citeasnoun{eivazi_pinns_pof}, although the PINNs approach is slower than a regular numerical integration of the governing equations, PINNs enable using coarser meshes due to the automatic differentiation. Therefore, the computational cost is similar to that of a numerical solver, with better performance when turbulence is present, {\it e.g.} for RANS simulations. A schematic representation of the proposed method is shown in \cref{fig:schematic}.

\subsubsection{Training:} Following the work by \citeasnoun{Raissi2019}, we employ a full-batch training procedure to learn function $f$. We initiate the training using the Adam optimizer \cite{kingma2017adam} for 1,000 epochs with the learning rate of $1\times10^{-3}$ and then apply the limited-memory Broyden--Fletcher--Goldfarb--Shanno (L-BFGS) algorithm \cite{Liu1989} to obtain the solution. The optimization process of the L-BFGS algorithm is stopped automatically based on the increment tolerance.

\section{Physics-informed super-resolution of flow fields in time and space}
\label{sec:SR}

We employ PINNs for super-resolution of flow-field data in time and space from only a limited set of low-resolution measurements containing noise and without having any high-resolution targets. The training data for supervised learning consists of only a few very low-resolution snapshots of the velocity components while we aim to obtain a continuous solution for velocity components and pressure. We consider three canonical test cases, \ie Burgers’ equation, two-dimensional vortex shedding behind a circular cylinder at $Re_D = 100$ \cite{Raissi2019} and a turbulent minimal channel at $Re_{\mathrm{cl}} = 5,000$ \cite{borrelli_predicting_2022} to show the applicability of PINNs for such super-resolution tasks in fluid mechanics. 
For all tests, we use fully-connected networks and consider the different weighting coefficients in the loss function based on the flow case. For Burgers’ equation, we utilized a neural network comprising 8 hidden layers with 20 neurons per hidden layer. For the cylinder wake and the minimal channel, we selected the model architecture based on the findings derived from hyper-parameter tuning process detailed in~\ref{sec:tune_vortex} and~\ref{sec:tune_channel}, respectively. For all the models, the hyperbolic tangent is used as the activation function. 

To evaluate the performance of the PINNs in the super-resolution tasks, we define the relative Euclidean norm of errors for a particular component as:
 {\begin{equation}\label{eq:rel}
    \epsilon_{\phi} = \left< \left\| \tilde{\phi} - \phi \right\|^2 / \left\| \phi \right\|^2 \right>, 
\end{equation}}
where $\tilde{\phi}$ represents the PINN simulation results and $\phi$ denotes the reference data for a particular component of the solution, which can be either one of the velocity components $(u, v, w)$ or the pressure $(p)$.~Note that $\|\cdot\|$ is the Euclidean norm and $\left<\cdot\right>$ is the average in the time. And, in order to quantify the prediction accuracy, we define the correlation coefficient ($r_{\phi}$) for a particular variable $\phi$ as:
 {\begin{equation}
    r_{\phi} = \frac{\sum_{i=1}^{n} (\tilde{\phi}_{i} - \overline{\tilde{\phi}})({\phi}_{i} - \overline{\phi})}{\sqrt{\sum_{i=1}^{n}(\tilde{\phi}_{i} - \overline{\tilde{\phi}})^2 \sum_{i=1}^{n}({\phi}_{i} - \overline{\phi})^2}}, 
    \label{eq:corrcoef}
\end{equation}}
where $\overline{\cdot}$ denotes the mean value of the quantities. The correlation coefficient is in the range of -1 to 1. Note that the higher linear correlation between prediction and reference is obtained, the closer the value to 1.

\subsection{Synthetic low-resolution data with noise}

In order to emulate possible measurement errors in an experiment, synthetic low-resolution data are generated from the reference velocity field by selecting a limited set of points in time and space. The low-resolution velocity field $\pmb{Q}_l$ is corrupted by an independent and identically-distributed (i.i.d) Gaussian noise as:
\begin{equation}
    \hat{\pmb{Q}}_l = \pmb{Q}_l \cdot (\mathbf{I} + c\pmb{\epsilon}),
\end{equation} 
where $\pmb{\epsilon} \sim  \mathcal{N}(\mathbf{0}, \mathbf{I})$ represents the noise factor, $c \in [0, 1]$ indicates the level of the noise and and $\mathbf{I}$ is the identity matrix.

\subsection{Burgers’ equation}

For the first test case, we consider the Burgers’ equation, which arises in various areas of applied mathematics, including fluid mechanics. We consider a one-dimensional case with Dirichlet boundary conditions and sinusoidal initial condition as:
\begin{equation}\label{eq:burgers}
    \begin{cases}
    \begin{aligned}
        &u_{t}+u u_{x}-(0.01 / \pi) u_{x x}=0, \quad x \in[-1,1], \quad t \in[0,0.99], \\
        &u(0, x)=-\sin (\pi x), \\
        &u(t,-1)=u(t, 1)=0.
    \end{aligned}
    \end{cases}
\end{equation}

\begin{figure}[ht]
    \centering
    \vspace{10pt}
    \begin{subfigure}{0.45\textwidth}
        (a)\par
        \includegraphics[height=0.7\textwidth]{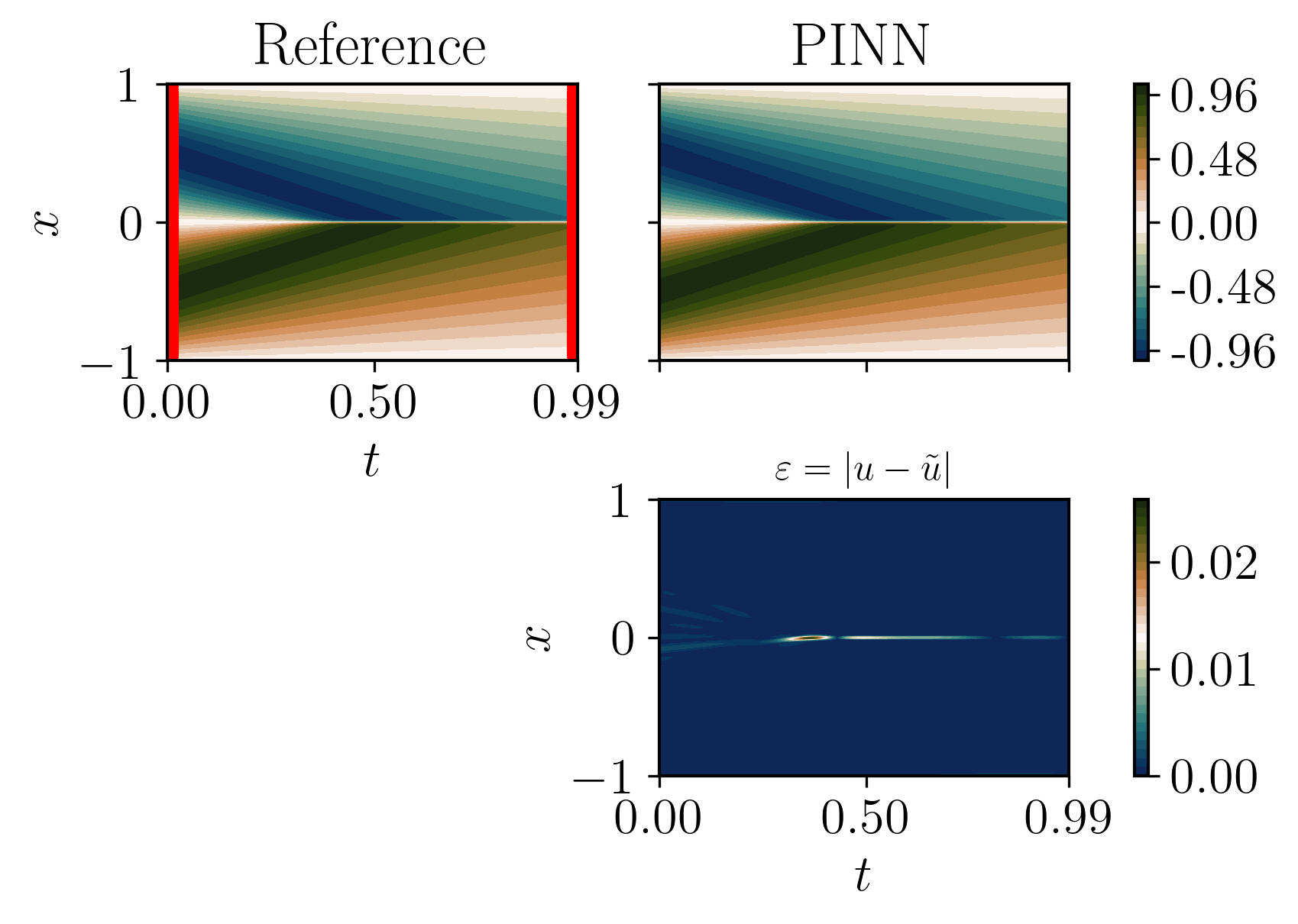}
    \end{subfigure}
    \begin{subfigure}{0.45\textwidth}
        (b)\par
        \includegraphics[height=0.7\textwidth]{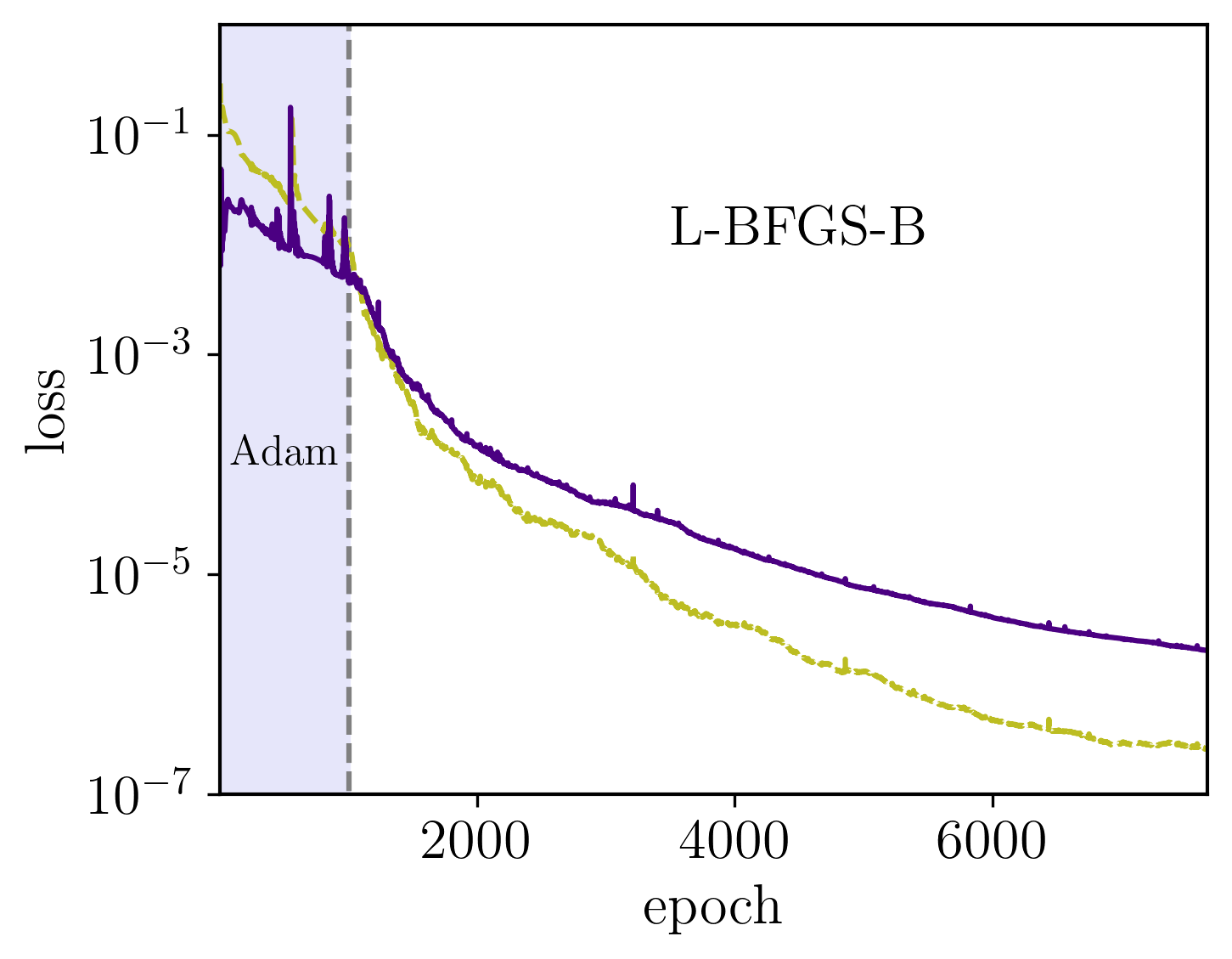}
    \end{subfigure}
    \caption{Super-resolution in time using PINNs for the one-dimensional Burger's equation: (a) comparison of the solution obtained from PINNs with the reference data; continuous solution of the shock-formation process (top) and the absolute error $\varepsilon = \left| u - \tilde{u} \right|$, where $u$ denotes the reference data and $\tilde{u}$ represents the PINN predictions (bottom). The red vertical lines indicate the data used for supervised learning. (b) Supervised and unsupervised losses, $L_s$ and $L_e$ respectively, during the training process. The green dashed line indicates $L_s$ and the blue line represents $L_e$.}
    \label{fig:burgers}
\end{figure}

Note that (\ref{eq:burgers}) is obtained from the Navier--Stokes equations for the velocity field by excluding the pressure-gradient term. We only take the data at the first time-step $t = 0$ and the last time-step $t = 0.99$ of the solution domain for supervised learning and our objective is to obtain a continuous solution in time. This can be seen as an extensive super-resolution task in time using only two snapshots of the data for training. Moreover, this is an ill-posed problem since we do not have the boundary conditions for the time-steps between $t = 0$ and $t = 0.99$. We consider $N_s = 512$ (256 for each time-step) and $N_e = 2,000$ randomly-generated collocation points for the training. Results are reported in \cref{fig:burgers}. The red vertical lines in \cref{fig:burgers}(a) show the data used for supervised learning. It can be observed that the PINN model can accurately provide a continuous solution of the shock formation process, see \cref{fig:burgers}(a), with a relative Euclidean norm of errors of $\epsilon_u = 0.13\%$ .~\Cref{fig:burgers}(b) shows the supervised and unsupervised losses, $L_s$ and $L_e$ respectively, during the training. At the final epoch of training the values $L_s = 2.59\times10^{-7}$ and $L_e = 2.04\times10^{-6}$ are obtained. These results show a good performance of PINNs in these super-resolution tasks in time, a fact that motivates their application to such problems in fluid mechanics.

\subsection{Two-dimensional vortex shedding behind a circular cylinder}
\label{sec:cylinder}
\begin{figure}[ht]
    \centering
    \includegraphics[width=0.8\textwidth]{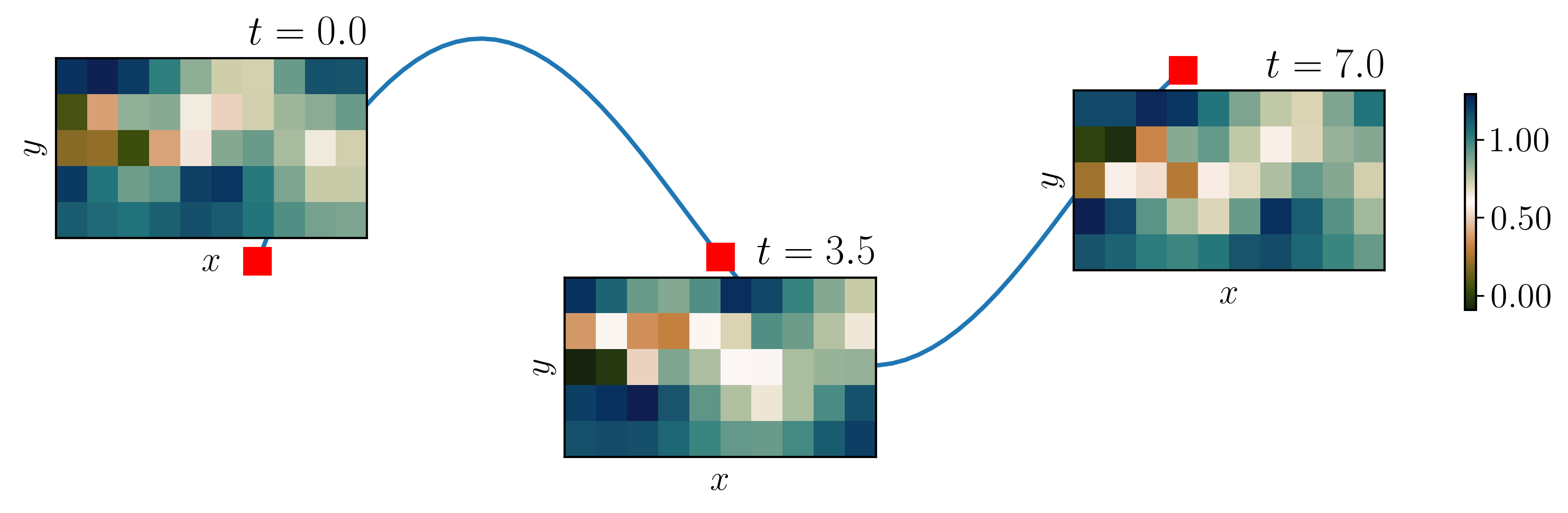}
    \caption{A schematic view of the resolution of the training data used for the supervised learning for the two-dimensional vortex shedding behind a circular cylinder. Colors depict the streamwise velocity. The blue line shows the first temporal coefficient $a_1$ of the POD for this test case.}
    \label{fig:cylinder_prob}
\end{figure}

For the second test case, we consider the super-resolution in time and space for the two-dimensional vortex shedding behind a circular cylinder at $Re_D = 100$. This example aims to highlight the ability of PINNs in the flow-field super-resolution tasks from very low-resolution measurements both in time and space. We consider the incompressible Navier--Stokes equations as:
\begin{equation} \label{eq:NS}
    \begin{aligned}
        &\frac{\partial \mathbf{u}}{\partial t}+(\mathbf{u} \cdot \nabla) \mathbf{u} =-\nabla p+\dfrac{1}{Re} \nabla^{2} \mathbf{u} \quad \text { in } \Omega, \\
        &\nabla \cdot \mathbf{u} =0 \quad \text { in } \Omega,
    \end{aligned}
\end{equation}
where $\mathbf{u}$ is the non-dimensional velocity vector, $t$ is the non-dimensional time and $p$ is the non-dimensional pressure. 
The reference data has a resolution of $(N_x, N_y, N_t) = (100, 50, 200)$ with a non-dimensional $\Delta t = 0.1$. The flow field is simulated by numerically integrating equation~(\ref{eq:NS}) over a time interval of almost one vortex-shedding cycle, \ie $t \in [0, 7]$. We perform downsampling on the velocity fields with a factor of 10 in each spatial direction and only utilize three snapshots of low-resolution data at $t = 0.0, 3.5$ and $7.0$ for supervised learning.~\Cref{fig:cylinder_prob} depicts the problem set-up for this test case and the reference data used for training. The blue line shows the first temporal coefficient $a_1$ of the proper orthogonal decomposition (POD) for this test case.

\begin{figure}[ht]
    \centering
    \begin{subfigure}{0.45\textwidth}
        (a)\par
    \end{subfigure}
    \begin{subfigure}{0.3\textwidth}
        (b)\par
    \end{subfigure}
    \includegraphics[width=0.8\textwidth]{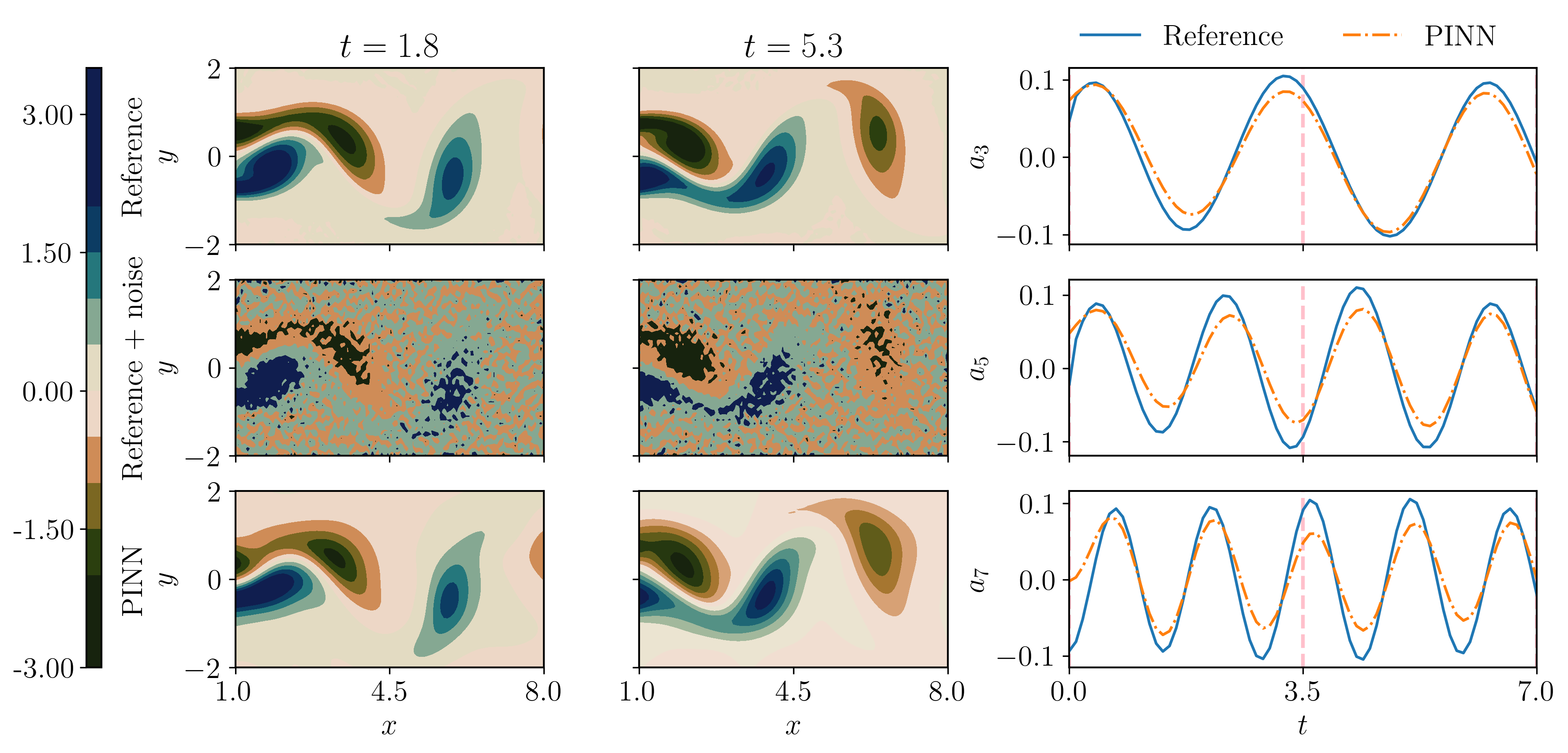}
    \caption{Results obtained based on super-resolution in time and space for the cylinder test case using PINNs from a limited set of measurements contaminated with a noise level of $c$ = 10\%: (a) contours of vorticity in the $z$ direction at $t$ = 1.8 and 5.3. (b) Temporal coefficients of the POD for the 3rd, 5th and 7th modes obtained from the PINN solution in comparison with that of the reference data.}
    \label{fig:cylinder_res}
\end{figure}

We utilize $N_s = 150$ points for supervised learning which is $0.14\%$ of the total amount of data, and consider $N_e = 2,000$ collocation points to compute the unsupervised loss. A hyper-parameter study of the PINNs architecture is shown in~\ref{sec:tune_vortex}, where the optimal architecture is reported. The PINN model contains 4 hidden layers with 20 neurons per hidden layer with the hyperbolic-tangent activation function whereas the weights for supervised and unsupervised learning are $\alpha$ = 1 and $\beta$ = 10, respectively. We assess the performance of PINNs in the reconstruction of the continuous solution from the low-resolution and noisy data. Noise levels of $c$ = 2.5\%, 5\% and 10\% are considered, and ~\cref{fig:cylinder_res} summarizes the results of our experiment for the noise level $c$ = 10\%. The middle panel shows the noisy data. It should be noted that these two snapshots are not used for supervised learning and the noisy data are depicted only for comparison. An adequate agreement can be observed between the solution obtained from PINNs and the reference data, as discussed below quantitatively. Moreover, it is of interest to assess the performance of PINNs in the reconstruction of the high-frequency modes during the cycle of vortex shedding from only three measurements in time. To this end, we project the solution predicted by the PINN model on the POD modes computed from 200 snapshots to obtain the temporal coefficients of the POD $a_i(t)$, where $i$ indicates the mode number.~\Cref{fig:cylinder_res}(b) shows the temporal coefficients of the POD for the 3rd, 5th and 7th modes obtained from the PINN solution in comparison with that of the reference data. The cross-correlation coefficients with respect to the reference temporal coefficients are 98.64, 96.54, and 94.37 for modes 3, 5, and 7, respectively, indicating that high-accuracy reconstruction of the temporal dynamics can be achieved even for high-frequency modes. Our results show the promising performance of the PINN model in enhancing the temporal resolution of the data and correcting the impact of experimental noise.

\subsubsection{Discussion of vortex-shedding results} 

\Cref{tab:errors} shows $\epsilon$ for the velocity components and pressure corresponding to the different noise levels. Despite the increasing noise levels leading to a rise in the relative Euclidean norm of errors $\epsilon$, the PINN model provides solutions with high accuracy. However, the errors of individual quantities do not explicitly represent the quality of the obtained reconstruction of vortex shedding via the proposed method. To address this, we evaluate the denoising results on the vorticity field $\omega$ using \cref{eq:corrcoef}. \Cref{tab:r_corr} reports the correlation coefficients obtained for the noisy reference and the predictions of PINNs with respect to the ground truth vorticity fields, respectively. The results indicate that PINNs exhibit promising performance in de-noising and reconstructing vorticity fields with limited measurements, highlighting the potential applications of our proposed method in experimental scenarios.

\begin{figure}[ht]
    \centering
    \includegraphics[width=0.6\textwidth]{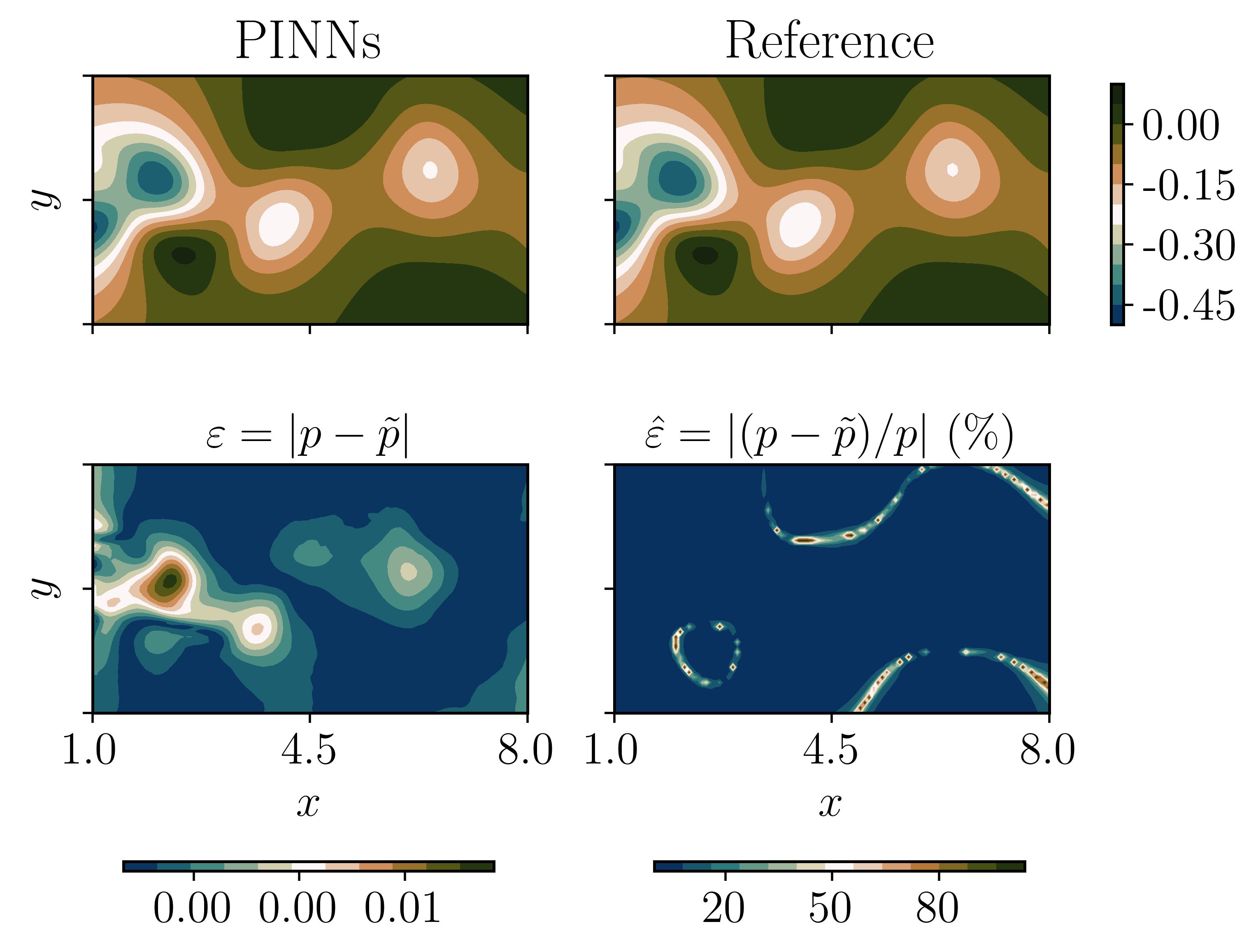}
    \caption{PINN-simulation results for the pressure at $t = 5.3$ for the clean test case (upper left), reference data (upper right), the absolute error $\varepsilon = \left| p - \tilde{p} \right|$ (lower left) and relative absolute error  {$\hat{\varepsilon} = |({p - \hat{p}})/{p}|$ $(\%)$} (lower right), where $p$ denotes the reference data and $\tilde{p}$ represents the PINN predictions.}
    \label{fig:cylinder_p}
\end{figure}

Note that to evaluate the accuracy of the PINN solutions, we apply a shift to the PINN-simulation results to bring the means of the pressure for reference data and the PINN results to the same value. We observe that the error for the pressure is higher than that of the velocity components since it is not used for supervised learning.~\Cref{fig:cylinder_p} depicts the PINN-simulation results for the pressure at $t = 5.3$ for the clean test case. It can be observed that although the absolute error $\varepsilon$ in some regions is slightly higher, the overall patterns in the flow are reconstructed with very good accuracy. The pressure at the peak of vorticity obtained by PINNs is $-0.237$, which meets the agreement with the ground truth of $-0.232$.

\begin{table}[ht]
    \centering
    \caption{Relative Euclidean norm of errors averaged over the whole domain and cross-correlation coefficient for the two-dimensional vortex shedding behind a circular cylinder.}
    \vspace{9pt}
    \begin{tabular}{ccccccc}
        \hline 
        \hline
        \vspace{5pt}
        
        Noise level & $\epsilon_{u}$ & $\epsilon_{v}$ & $\epsilon_{p}$& $r_{u}$ & $r_{v}$ & $r_{p}$ \\
        \hline
        
        \rule{0pt}{3ex}clean & 2.47\% & 5.92\% & 14.32\% & 0.997 & 0.998 & 0.979 \\
        2.5\% & 3.13\% & 6.92\% & 10.63\% & 0.995 & 0.997 & 0.990\\
        5\% & 3.96\% & 8.66\% & 19.06\% & 0.992 & 0.996 & 0.961 \\
        10\% & 7.24\% & 13.12\% & 33.07\% & 0.973 & 0.991 & 0.890 \\
    \hline
    \hline
    \end{tabular}
    \label{tab:errors}
\end{table}

\begin{table}[ht]
    \centering
    \caption{Cross-correlation coefficient for vorticity field $\omega$ obtained for each noise level  {in} the noisy reference data and the prediction of PINNs. For each noise level, we denote the highest correlation in bold.}
    \vspace{9pt}
    \begin{tabular}{ccc}
        \hline 
        \hline
        \vspace{5pt}
        Type & Noise level & $r_{\omega}$ \\
        \hline
        \rule{0pt}{3ex}Noisy reference & $2.5\%$ & 0.948 \\
        PINNs & $2.5\%$ & 0.985 \\ \hline
        \rule{0pt}{3ex}Noisy reference & $5\%$ & 0.831 \\ 
        PINNs & $5\%$ & 0.981 \\  \hline
        \rule{0pt}{3ex}Noisy reference & $10\%$ & 0.598 \\ 
        PINNs & $10\%$ & 0.954 \\  \hline
    \hline
    \hline
    \end{tabular}
    \label{tab:r_corr}
\end{table}

\subsection{Minimal turbulent channel}
\label{sec:channel}
We employ a dataset~\cite{borrelli_predicting_2022} of the minimal turbulent channel at $Re_{\mathrm{cl}} = 5,000$ generated following the parameters described in the work by~\citeasnoun{jimenez_moin}. This database was obtained through direct numerical simulation as discussed by \citeasnoun{borrelli_predicting_2022}. The box size was set to $x_l = 0.6 \pi h$, $y_l = 2h$ and $z_l = 0.18 \pi h$, where $h$ denotes the half-height of the channel. The resolution is $(N_x, N_y, N_z) =  (32, 129, 16)$, representing the number of grid points in the streamwise ($x$), wall-normal ($y$) and spanwise ($z$) directions, respectively. We take a time interval of $t = [0, 4]$ and a domain size of $x_d = 0.6 \pi h$, $y_d = h$ and $z_d = 0.01125 \pi h$ for our super-resolution task. We assess the performance of PINNs in the reconstruction of a continuous solution from a limited set of measurements with different resolutions in time and space and the presence of noise. For this test case, we employ a PINN model with 10 hidden layers, each containing 100 neurons with hyperbolic tangent as the activation function and the three-dimensional incompressible Navier--Stokes equations, the results of hyper-parameter studies are reported in~\ref{sec:tune_channel}. For time, we consider two temporal resolutions: $\Delta t = 1.0$ and 2.0 corresponding to, respectively, taking only five and three low-resolution snapshots for supervised learning. And, we utilize two different resolutions in space, \ie $(n_x, n_y, n_z) = (8, 8, 2)$ and $(16, 16, 2)$ where $n_x, n_y$ and $n_z$ represent the number of selected points in the $x$, $y$, and $z$ directions, respectively. The spacing between the selected points in the $x$ and $z$ directions is constant, but the spacing in the $y$ direction $\delta y$ is not uniform: the closer to the wall the lower $\delta y$. We represent different datasets with different resolutions as t\#--s\#, which indicate the number of selected points in time and space.

\begin{figure}[ht]
    \centering
    \includegraphics[width=0.8\textwidth]{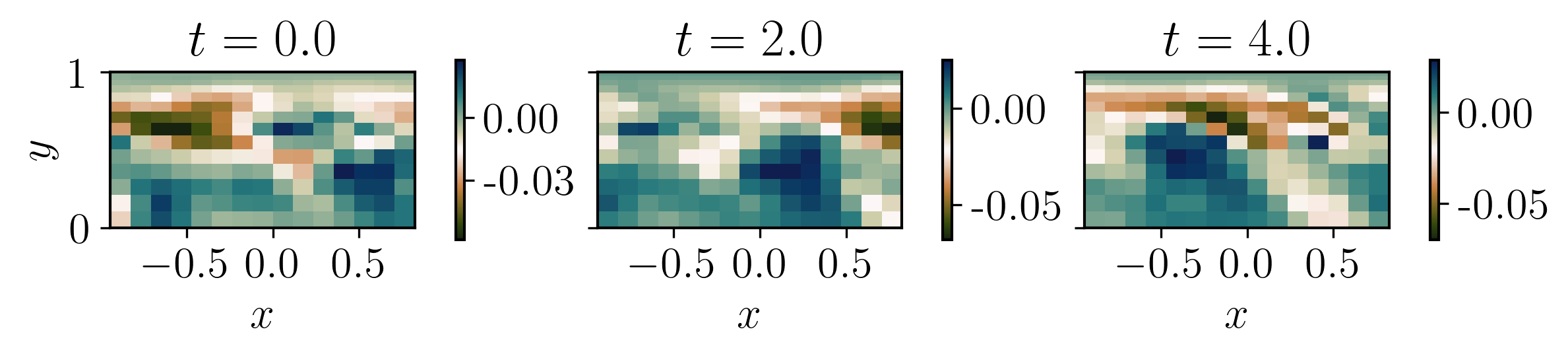}
    \caption{Training dataset t3--s16 for the supervised learning of the minimal turbulent channel. Colors represent the instantaneous wall-normal velocity $v$.}
    \label{fig:channel_data}
\end{figure}

The trained PINN model with each dataset is represented as PINN--t\#--s\#.~\Cref{fig:channel_data} shows the instantaneous wall-normal velocity $v$ for the dataset t3--s16, which has a low resolution in time and a moderate resolution in space.
As the minimal turbulent channel is focused on near-wall turbulence~\cite{jimenez1991minimal}, we scale the selected spatial and temporal resolutions by the viscous length $l^{*} = \nu / u_{\tau}$ (where $\nu$ is the fluid kinematic viscosity, and $u_{\tau} = \sqrt{\tau_w / \rho}$ is the friction velocity) and viscous time $t^* = \nu / u^2_{\tau}$. \Cref{tab:scale_channel} summarizes the inner-scaled time interval $\Delta t/t^*$, inner-scaled spacings in the $x$ and $z$ directions ($\delta_x/l^*$, $\delta_z/l^*$), as well as the inner-scaled minimum and maximum spacing in the $y$ direction ($\min \delta_y/l^*$, $\max \delta_y/l^*$). The scaling results indicate that the adopted lower resolutions drastically decrease the temporal and spatial resolutions, sampling around 1$\%$ of the total data as the ground truth for supervised learning.

\begin{table}[ht]
    \centering
    \caption{Spatial and temporal separations of utilized resolutions scaled by viscous length ($l^*$) and viscous time ($t^*$) in the minimal turbulent channel.}
    \vspace{9pt}
    \begin{tabular}{cccccc}
        \hline 
        \hline
        \vspace{5pt}
        
        Case & $\delta t/t^*$ &  $\delta_x/l^*$ & $\min \delta_y/l^*$ & $\max \delta_y/l^*$ & $\delta_z/l^*$\\
        \hline

        \rule{0pt}{3ex}DNS & 1.63 & 11.90 & 0.06 & 4.95 & 7.14  \\
        \rule{0pt}{3ex}PINN--t5--s16 & 8.16  & 23.80 & 0.97 & 19.61 & 7.14 \\ 
                    PINN--t3--s16 & 16.32 & 23.80 & 0.97 & 19.61 & 7.14 \\ 
        \rule{0pt}{3ex}PINN--t5--s8 & 8.16  & 47.60 & 1.94 & 39.22 & 7.14 \\
                        PINN--t3--s8& 16.32 & 47.60  & 1.94 & 39.22 & 7.14\\

    \hline
    \hline
    \end{tabular}
    \label{tab:scale_channel}
\end{table}

\begin{table}[ht]
    \centering
    \caption{Relative Euclidean norm of errors averaged over the whole domain and correlation coefficient for the minimal turbulent channel.}
    \vspace{9pt}
    \begin{tabular}{cccccccc}
        \hline 
        \hline
        \vspace{5pt}
        
        Test & Noise level & $\epsilon_{u}$ & $\epsilon_{v}$ & $\epsilon_{w}$ & $r_{u}$ & $r_{v}$ & $r_{w}$\\
        \hline
        
        \rule{0pt}{3ex}PINN--t5--s16 & clean & 0.53\% & 12.54\% & 9.89\% & 0.999 & 0.991 & 0.995\\
        PINN--t3--s16 & clean & 0.81\% & 17.57\% &  20.75\%  & 0.999 & 0.9807 & 0.9871\\ 
        
        \rule{0pt}{3ex}PINN--t5--s8 & clean & 1.87\% & 22.86\% &  44.15\% & 0.999 & 0.9608 & 0.9039\\
        PINN--t3--s8 & clean & 2.64\% & 46.72\% &  56.60\%  & 0.998 & 0.8655 & 0.8563\\
        
        \rule{0pt}{3ex}PINN--t5--s16--c5.0 & 5.0\% & 4.61\% & 21.78\% & 19.19\% & 0.9973 & 0.9749 & 0.9801\\
        PINN--t3--s16--c5.0 & 5.0\% & 5.40\% & 36.74\% & 35.11\% & 0.9959 & 0.8792 & 0.9430 \\
        
    \hline
    \hline
    \end{tabular}
    \label{tab:errors_channel}
\end{table}

\Cref{tab:errors_channel} summarizes the results of our experiments for the minimal turbulent channel using the training data of different resolutions and for different noise levels. Our results show great potential from the PINN model by using the t5--s16 dataset for supervised learning, leading to $\epsilon_u$ = 0.53\%, $\epsilon_v$ = 12.54\% and $\epsilon_w$ = 9.89\% whereas $r_u$ = 0.999, $r_v$ = 0.991 and $r_w$ = 0.995. An adequate performance can also be achieved using the dataset t3--s16, which leads to $\epsilon_u$ = 0.81\%, $\epsilon_v$ = 17.57\% and $\epsilon_w$ = 20.75\% whereas $r_u$ = 0.999, $r_v$ = 0.981 and $r_w$ = 0.987. Note that the presented trend shows the evolution of the error with increasing downsampling, and although in an experiment with no ground-truth data the actual error would not be possible to quantify, it is possible to establish estimates of expected errors based on numerical trends. Our results show the good performance of PINNs for super-resolution of turbulent flows both in time and space from only a limited set of measurements and without having any high-resolution target data. This is a very relevant result in the context of data augmentation for experiments in fluid mechanics.

\begin{figure}[ht]
    \centering
    \includegraphics[width=0.8\textwidth]{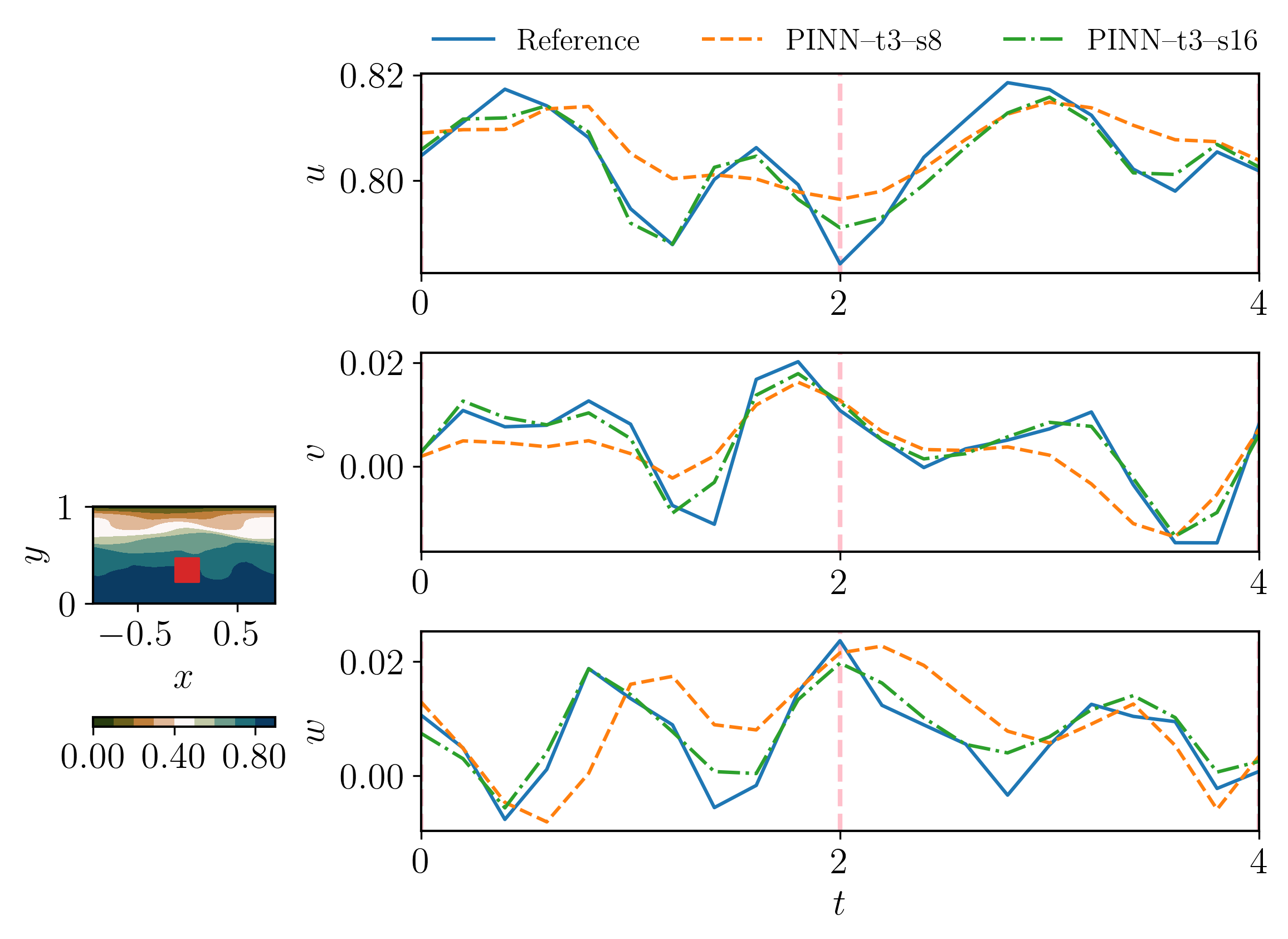}
    \caption{The evolution of the velocity components in time for a point at $(x, y) = (0.0, 0.36)$ (marked by the red square) obtained from the PINN--t3--s16 and PINN--t3--s8 models in comparison with the reference data. Contour represents the instantaneous spanwise velocity at $t$ = 0.0.}
    \label{fig:channel_sigs}
\end{figure}

Further reduction of the spatial resolution leads to a considerable increase in error. For instance, the PINN model PINN--t3--s8 yields simulation results with $\epsilon_u$ = 2.64\%, $\epsilon_v$ = 46.72\%, and $\epsilon_w$ = 56.60\%, with corresponding correlation coefficients of $r_u$ = 0.998, $r_v$ = 0.8655, and $r_w$ = 0.8563. This may be associated with the fact that the magnitudes of $v$ and $w$ are lower than that of $u$, thus being more sensitive to the small variation when computing the error via \cref{eq:rel}. However, it is noteworthy that the obtained $r_v$ and $r_w$ indicate that the wall-normal and spanwise velocity components are reconstructed with adequate accuracy.

We compare the performance of the PINN models PINN--t3--s16 and PINN--t3--s8: \Cref{fig:channel_sigs} illustrates the evolution of the velocity components in time for a point at $(x, y) = (0.0, 0.36)$, using both of the aforementioned models. The red square shows the location of the point on the contour of the streamwise velocity $u$. For these cases, we only use the data at $t$ = 0.0, 2.0, and 4.0 for supervised learning. It can be observed that although the training data for supervised learning is sparse, the PINN--t3--s16 model provides a very good reconstruction of the velocity components in time, yielding cross-correlation coefficients of 0.949, 0.965, and 0.947 for the streamwise, wall-normal, and spanwise velocities, respectively. However, the results obtained from the PINN--t3--s8 model are not as accurate, with cross-correlation coefficients of 0.791, 0.776, and 0.584 for $u$, $v$, and $w$, respectively.

\subsubsection{Discussion of minimal-channel results}

\begin{figure}[ht]
    \centering
    \includegraphics[width=0.8\textwidth]{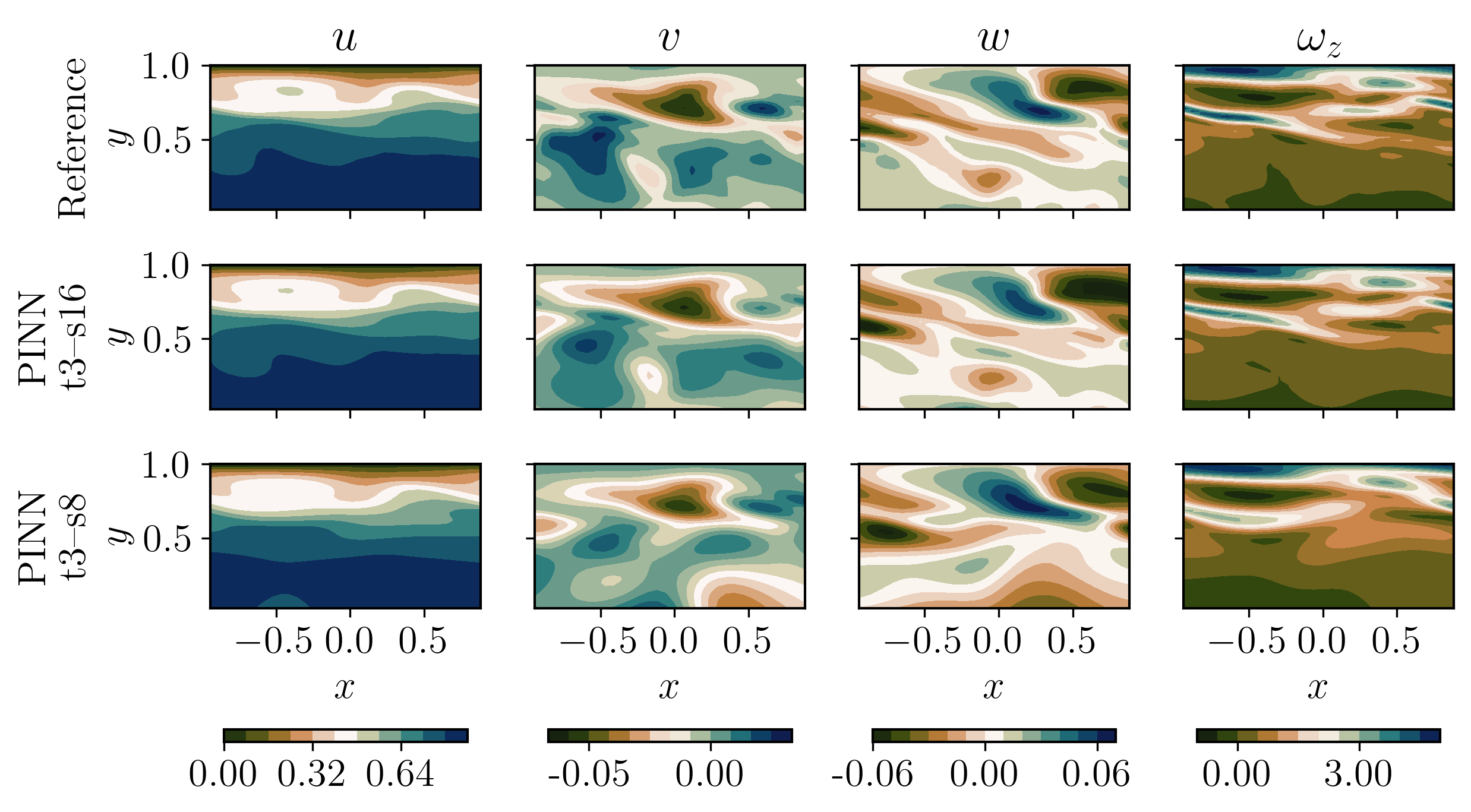}
    \caption{Contours of the instantaneous velocity components $(u, v, w)$ and the spanwise vorticity $(\omega_z)$ obtained from the PINN--t3--s16 and the PINN--t3--s8 models compared with the reference data at $t$ = 1.0.}
    \label{fig:channel_r1}
\end{figure}

\begin{figure}[ht]
    \centering
    \includegraphics[width=0.8\textwidth]{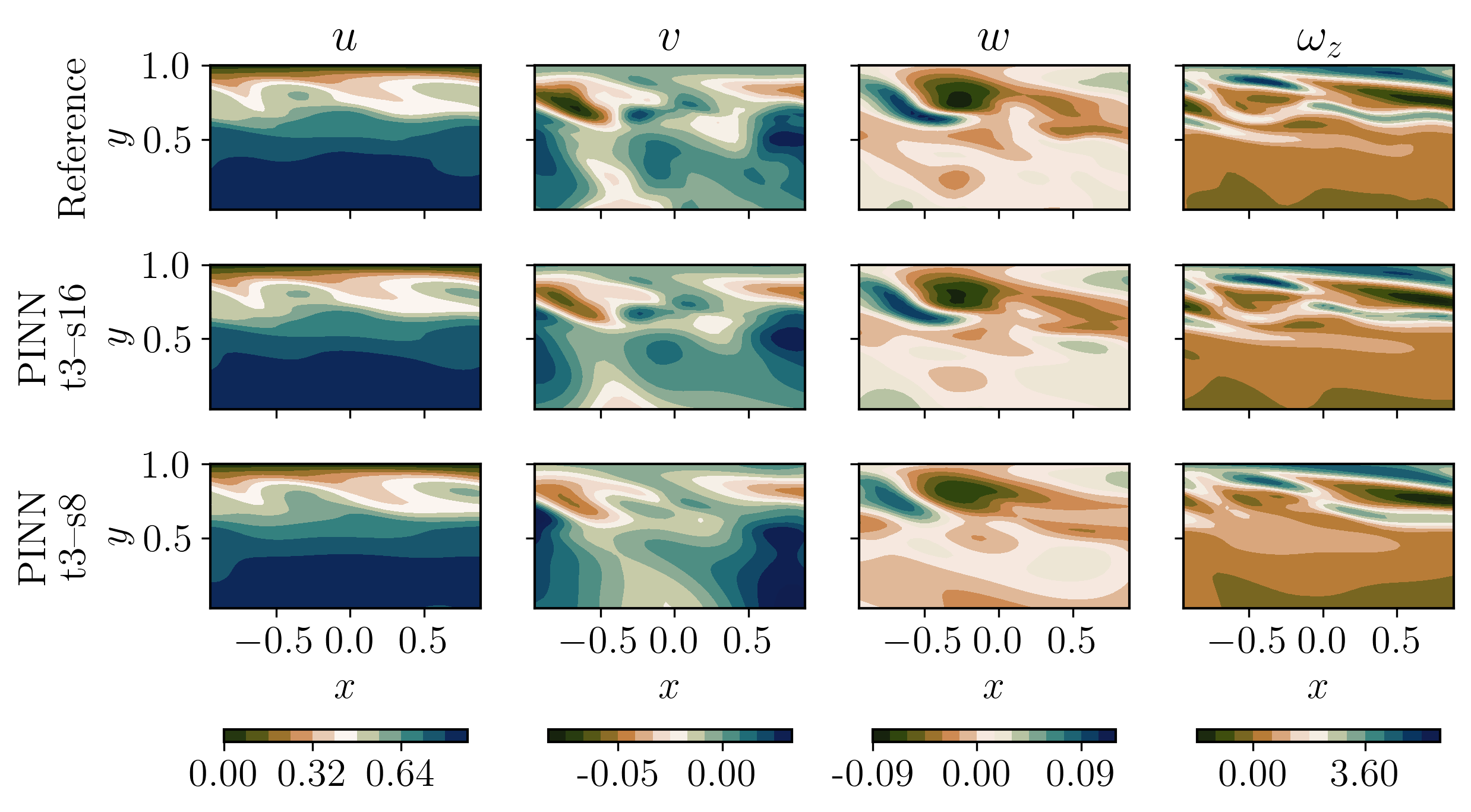}
    \caption{Contours of the instantaneous velocity components $(u, v, w)$ and the spanwise vorticity $(\omega_z)$ obtained from the PINN--t3--s16 and the PINN--t3--s8 models compared with the reference data at $t$ = 3.0.}
    \label{fig:channel_r2}
\end{figure}

\Cref{fig:channel_r1,fig:channel_r2} depict contours of the velocity components $(u, v, w)$ and the spanwise vorticity $(\omega_z)$ obtained from the PINN--t3--s16 and the PINN--t3--s8 models compared with the reference data at $t$ = 1.0 and 3.0, respectively. Our results demonstrate that the PINN model PINN--t3--s16 can effectively reconstruct the continuous solution of the velocity components in all three directions, with cross-correlation coefficients yielding $r_u$ = 0.999, $r_v$ = 0.981, and $r_w$ = 0.9871, indicating promising accuracy. Additionally, the relative Euclidean errors of the wall-normal and spanwise velocities are observed to be much higher than that of the streamwise velocity. 

\begin{figure}[ht]
    \centering
    \begin{subfigure}{0.45\textwidth}
        \centering
        \includegraphics[width=0.78\textwidth]{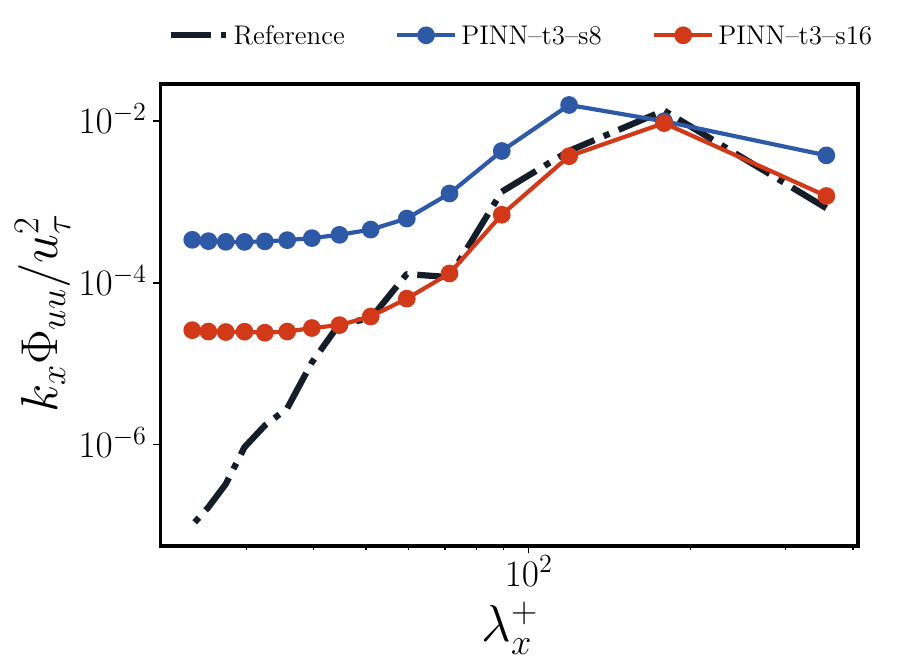}
        \caption{Streamwise velocity ($u$)}
    \end{subfigure}
    \quad
    \begin{subfigure}{0.45\textwidth}
        \centering
        \includegraphics[width=0.8\textwidth]{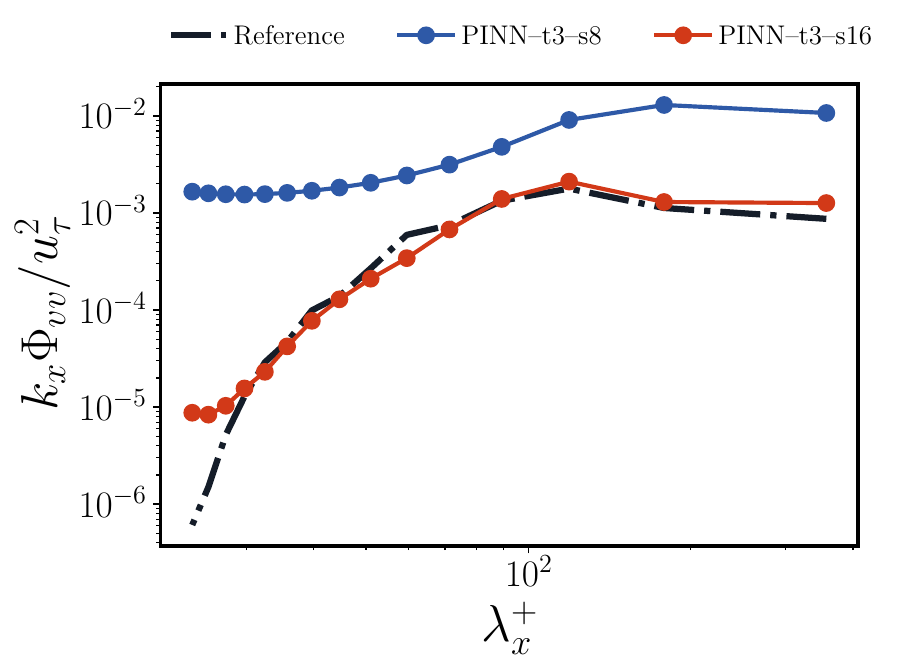}
        \caption{Wall-normal velocity ($v$)}
    \end{subfigure}
    \quad
    \begin{subfigure}{0.45\textwidth}
        \centering
        \includegraphics[width=0.8\textwidth]{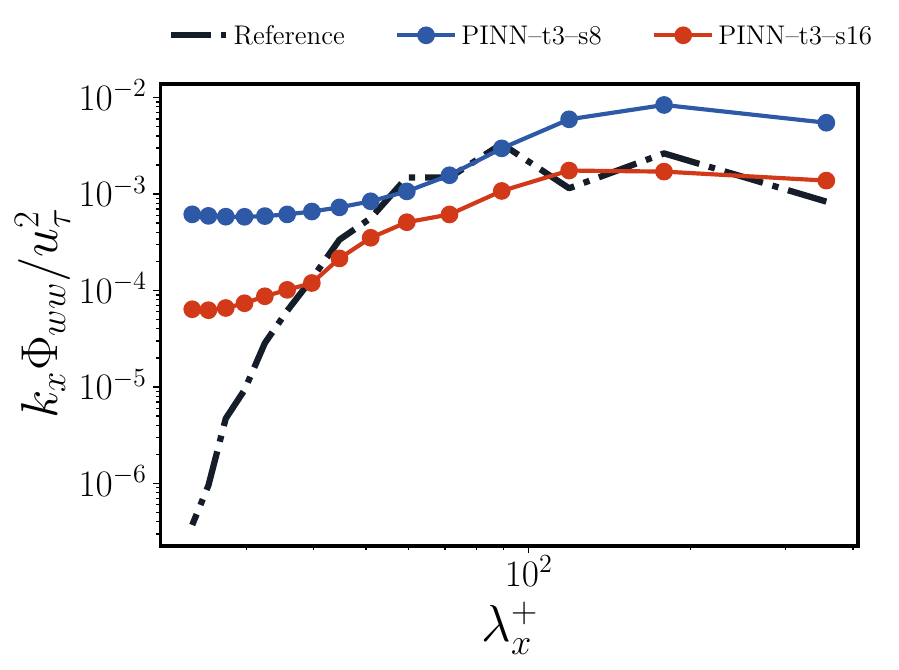}
        \caption{Spanwise velocity ($w$)}
    \end{subfigure}
    \caption{Streamwise one-dimensional premultiplied power-spectral density (PSD) of the streamwise velocity ($u$), wall-normal velocity ($v$) and spanwise velocity ($w$) components at $y^+ = 30$.}
    \label{fig:psd_channel}
\end{figure}

Moreover, it is of interest to investigate the range of scales that can be reconstructed by the various models. To this end, we computed the inner-scaled streamwise one-dimensional premultiplied power-spectral density (PSD) of the various velocity components at $y^+ = 30$ obtained by the PINN--t3--s16 and PINN--t3--s8 models, as depicted in \cref{fig:psd_channel}. The reconstructed spectra obtained by PINN--t3--s16 agrees very well with the reference data for the larger wavelengths, indicating that the model can reconstruct the large-scale features of the flow. For the streamwise, wall-normal and spanwise velocity components, the spectra exhibit deviations with respect to the ground truth for the scales smaller than $\lambda^+_x = 51.0$, 39.7 and 32.4, respectively. Furthermore, the spectra reconstructed by PINN--t3--s8 exhibit higher deviations with respect to the ground truth with those obtained by PINN--t3--s16 model for all the velocity components, a fact that is caused by the reduced spatial resolution.

The effect of noise on the accuracy of the reconstructions is also assessed and the results are reported in \cref{tab:errors_channel}. We observe that very good results can be achieved, even in the presence of 5\% noise, from the PINN--t5--s16 and PINN--t3--s16 models. For the PINN--t5--s16 model trained on the noisy data, the highest error is obtained for the wall-normal velocity and it is equal to $\epsilon_w$ = 21.78\%, while the lowest error is again obtained for the streamwise velocity, and it is equal to $\epsilon_u$ = 4.61\%. The cross-correlation coefficient of spanwise vorticity obtained by the PINN model PINN--t3--s16--c5 is 0.98, while PINN--t5--s16--c5 yields 0.94. In comparison, the noisy reference data yields $r_{\omega_z}$ = 0.81, indicating the promising performance of PINNs in removing noise from the reference data.

\section{ Application to experimental dataset} \label{sec:exp}

\begin{figure}[ht]
    \centering
    \includegraphics[width=0.5\textwidth]{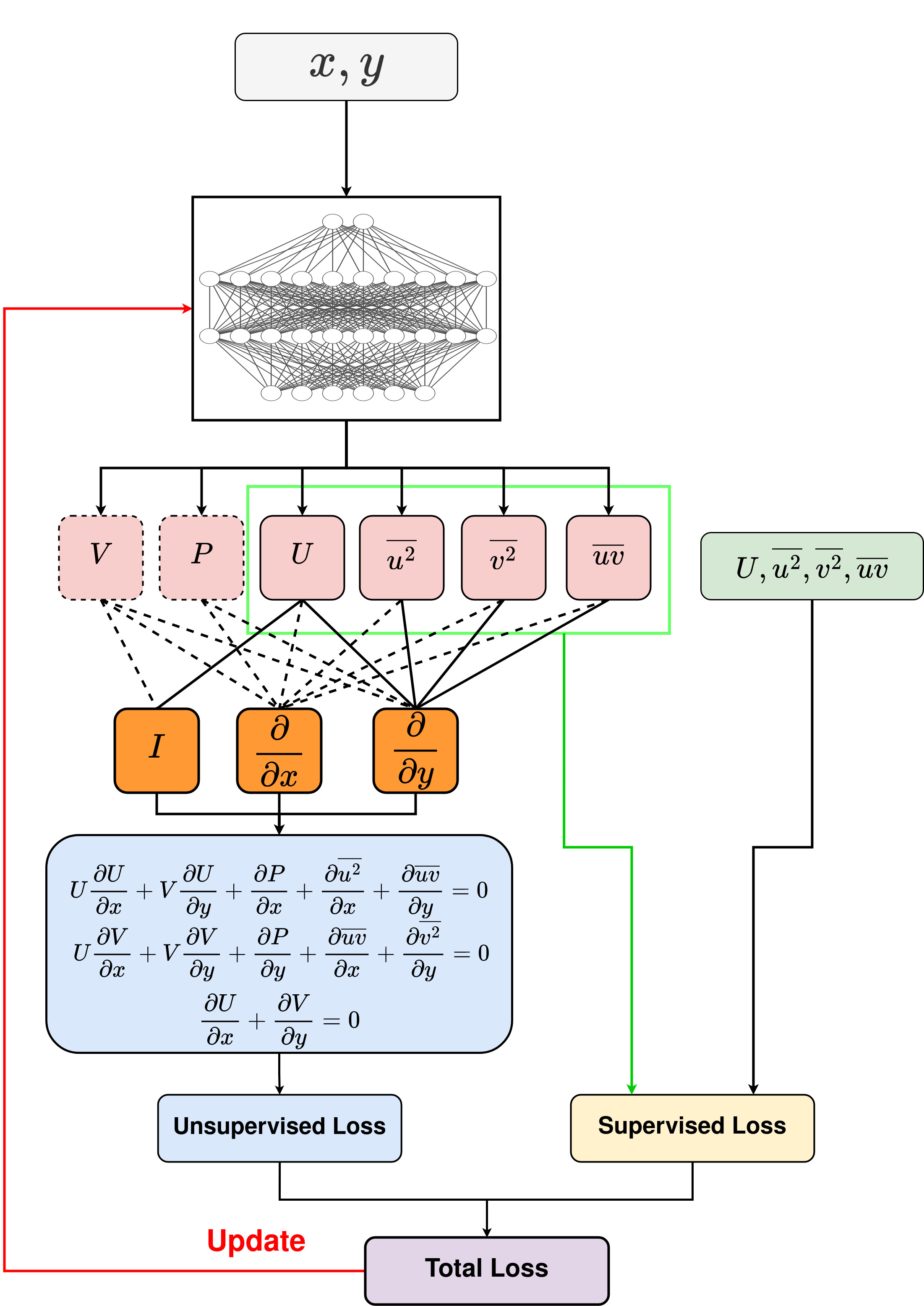}
    \caption{ Schematic representation of the PINNs approach employed here. The inputs are the coordinates, and the neural network predicts the various profiles. These profiles are used, through automatic differentiation, to construct the 2D RANS equations without viscous terms, yielding the unsupervised loss. Furthermore, some of the outputs of the network are compared with the known reference values, yielding the supervised loss. These two are combined to obtain the total loss, and the training process is continued.}
    \label{fig:pinns_schematic}
\end{figure}

In this section, we will discuss the application of PINNs to an experimental dataset consisting of turbulence measurements conducted by hot-wire anemometry~\cite{bayoan_et_al}. This dataset includes measurements of the mean and the Reynolds stresses in a rough turbulent boundary layer which constitutes the inflow for an array of wind turbines. We will introduce two different synthetic errors in this data, namely low resolution in the wall-normal direction and noisy measurements, and we will assess the capabilities of the PINNs approach to correct them. The interest of this particular dataset lies in the fact that it lacks information regarding the mean wall-normal velocity $V$ and pressure $P$, a fact that might complicate the predictions. The particular PINNs framework employed here is described in Figure~\ref{fig:pinns_schematic}, where it can be observed that we employed the 2D RANS equations without viscous terms. This is in agreement with the original experimental study~\cite{bayoan_et_al}, where it is claimed that in this flow the most relevant physics takes place far from the wall. The results for both synthetic errors are presented next.

\subsection{ Improve resolution}\label{sec:super_resolution}

 We begin by testing the capabilities of PINNs to enhance resolution in a one-dimensional profile by removing 6, 8, and 10 of the 13 points in the wall-normal direction of the initial profiles, respectively. To evaluate how the proposed PINN approach compares to the state-of-the-art methods, we employ spline interpolation and a neural network (NN) using the same architecture as PINNs to enhance resolution.
 
\begin{figure}[ht]
    \centering
    \begin{subfigure}{\textwidth}
        \centering
        \includegraphics[width=\textwidth]{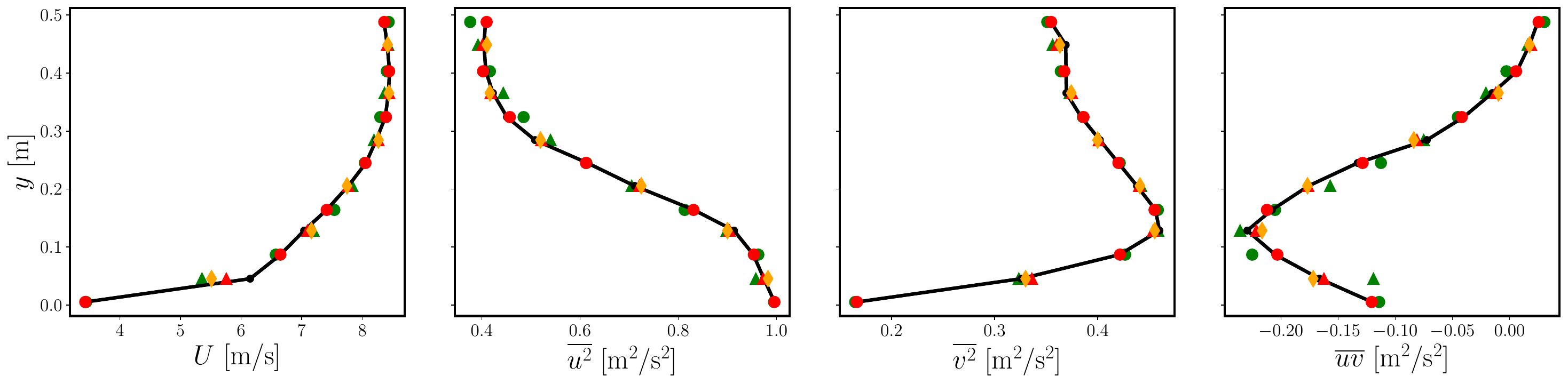}    
    \end{subfigure}
    \quad
    \begin{subfigure}{\textwidth}
        \centering
        \includegraphics[width=\textwidth]{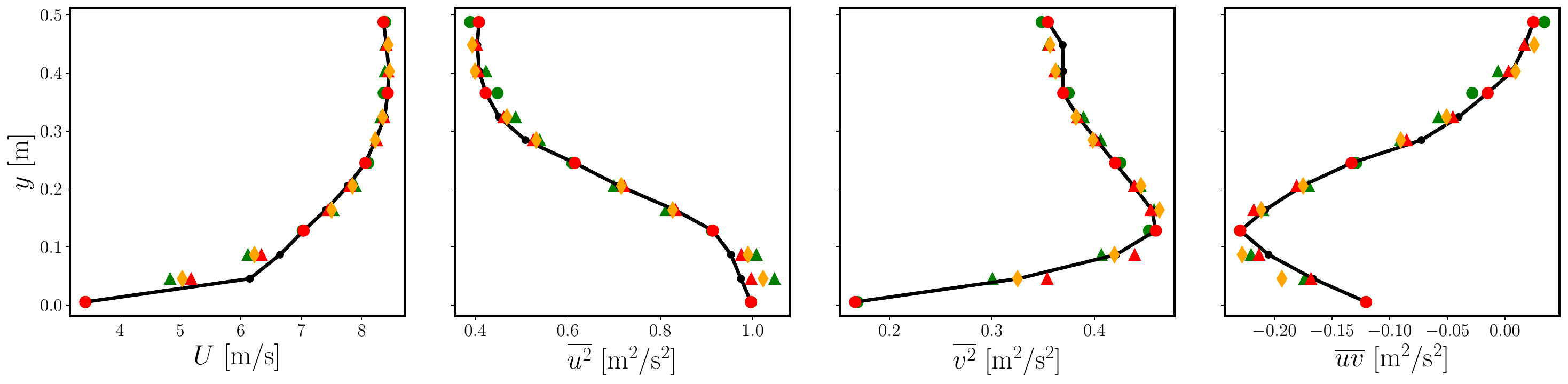}    
    \end{subfigure}
    \quad
    \begin{subfigure}{\textwidth}
        \centering
        \includegraphics[width=\textwidth]{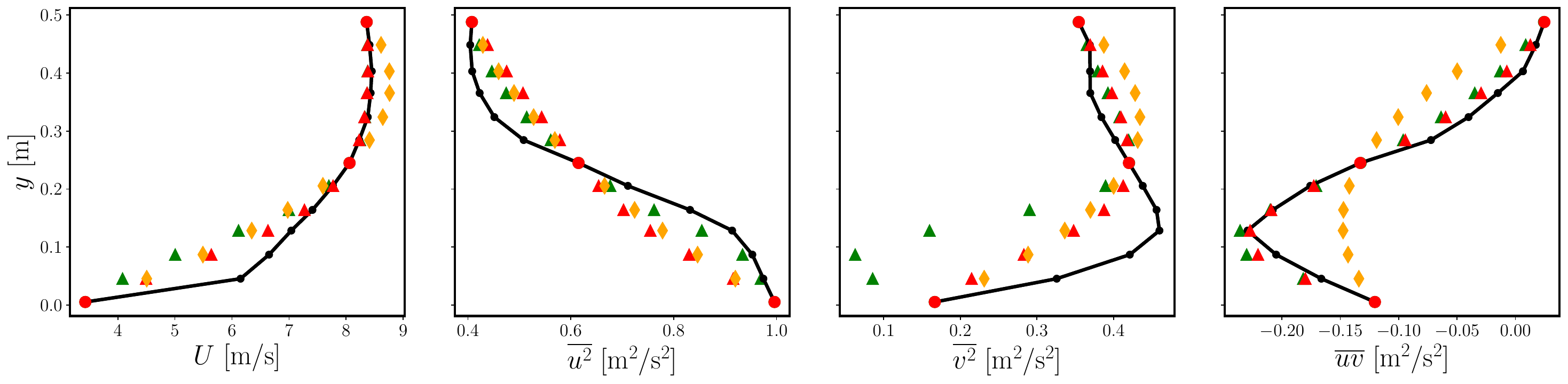}    
    \end{subfigure}
    \quad
    \caption{Super-resolution results on (from left to right) mean flow, streamwise fluctuations, wall-normal fluctuations and Reynolds shear stress. (Black) original profile, (red) results obtained by PINNs, (yellow) results obtained by spline interpolation and (green) results obtained by NNs. Note that we denote the remaining points after reducing resolution as circle and added points as triangle markers, respectively. From the top to the bottom we show the results obtained retaining 7, 5 and 3 reference points, respectively.}
    \label{fig:hotwire_sr}
\end{figure}

 As illustrated in \cref{fig:hotwire_sr}, the retained 7, 5, and 3 points (marked as circles) will be used for the supervised part of the training, while the unsupervised part will involve minimizing the residual of the two-dimensional RANS equations (without viscosity) on 50 randomly-sampled points. A hyper-parameter study of the PINNs architecture is presented in~\ref{sec:tune_hotwire}, where the optimal architecture is reported. Table~\ref{tab:hotwire_sr} summarizes the relative errors produced by the three considered methods using different numbers of reference points for enhancing resolution. As observed in figure~\ref{fig:hotwire_sr}, when retaining 7 and 5 points, the newly-added points (marked as triangles) are in excellent agreement with the reference profile (in black). The averaged relative Euclidean norm of error on added data are $\overline{\epsilon}$ of 2.16$\%$ and 3.76$\%$ as retaining 7 and 5 points, respectively, highlighting the potential of PINNs for super-resolution tasks. Our results demonstrate that the proposed PINNs achieve the lowest average error $\overline{\epsilon}$ among the considered methods in all cases, highlighting the good performance of PINNs for super-resolution of turbulent measurements. It is worth noting that even when only using 3 points for training, the average error obtained by PINNs is still around $10.98\%$, whereas for spline interpolation, it rises to $17.84\%$, demonstrating the robustness of our proposed methods.

\begin{table}[ht]
    \centering
    \caption{Relative Euclidean norm of errors using different number of retained reference points for enhancing the resolution. The boldface denotes the lowest averaged error among the obtained results in each case.}
    \begin{tabular}{ccccccc}
                \hline
                \hline 
              Methods & Ref. Points &  $\epsilon_{U}$ ($\%$) & $\epsilon_{\overline{u^2}}$ ($\%$) & $\epsilon_{\overline{v^2}}$ ($\%$) & $\epsilon_{\overline{uv}}$ $\%$
 & $\overline{\epsilon}$ ($\%$) \\\hline
                \rule{0pt}{3ex}PINNs   &   7     &   {1.46}   & {0.84}        & 1.16 & {2.98} & {1.61} \\
                NNs     &   7      &   3.07    & 2.72          & 1.19   & 12.46 & 4.86 \\
                Spline  &   7     &   2.38   & 0.99        & {0.74}  & 3.90 &  2.00 \\ \hline
                \rule{0pt}{3ex}PINNs   &  5        &   {3.73}   & {1.54}        & 2.65 & {4.06} & {2.99} \\ 
                NNs     &  5       &   5.26    & 4.38          & 2.65   & 7.65 & 4.98 \\ 
                Spline   &  5     &   4.40   & 2.72        & {1.31}  & 8.66 & 4.27  \\ \hline
                \rule{0pt}{3ex}PINNs    &    3   &   {7.29}   & {11.86}        & {16.25} & {8.55} & {10.98} \\ 
                NNs      &    3    &   10.38    & {5.74}          & 39.79   & 11.10  & 16.75\\ 
                Spline   &   3    &   8.27   & 9.96        & {17.56}  & 35.55 & 17.84 \\ 
                \hline 
                \hline
            \end{tabular}
    
    \label{tab:hotwire_sr}
\end{table}

\subsubsection{Discussion of super-resolution results}

\begin{figure}[ht]
    \centering
    \includegraphics[width=0.45\textwidth]{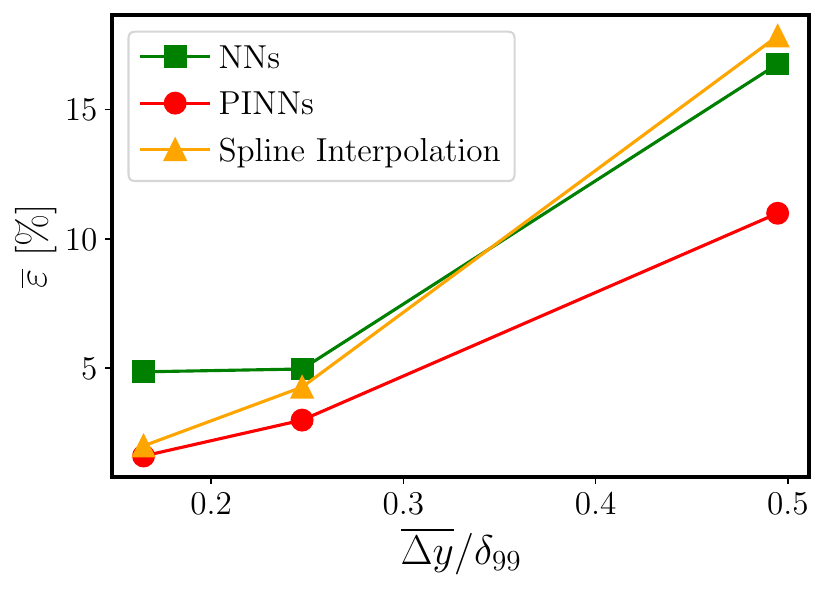}
    \caption{Averaged relative Euclidean norm of the errors as a function of the averaged spacing between retained samples ($\overline{\Delta y}$) normalized by the boundary layer thickness ($\delta_{99}$).}
    \label{fig:hotwire_sr_dist_vs_error}
\end{figure}

Our results indicate that PINNs can effectively improve the resolution of the original velocity profiles, achieving average relative errors of $1.61\%$, $2.99\%$, and $10.98\%$ using 7, 5, and 3, respectively. Figure~\ref{fig:hotwire_sr_dist_vs_error} shows the averaged relative Euclidean norm of the errors as a function of the spacing between retained samples normalized by the boundary layer thickness ($\delta_{99}$). It can be observed that the error drastically rises with increasing spacing between retained points, while the error obtained by PINNs exhibits the lowest variation with respect to the spacing between retaining points compared with spline interpolation and NNs. It is also interesting to note that when we used 7 reference points, the vanilla NNs perform worse than spline interpolation, highlighting the advantages of incorporating physical information into neural networks. As depicted in \cref{fig:hotwire_sr}, the most challenging quantity to predict is the Reynolds shear stress. This difficulty can be attributed to the inherent complexities involved in its measurement, particularly in the near-wall region. Additionally, the relative error in predicting $\overline{uv}$ is significantly higher compared with other quantities, primarily due to its lower absolute values, rendering it highly sensitive to small variations. Note that the error incurred through spline interpolation is more than double and quadruple that of PINNs when retaining 5 and 3 points, respectively. This observation underscores the potential of PINNs-based approaches in enhancing the resolution of experimental measurements.

\subsection{De-noising}
\begin{figure}[ht]
    \centering
    \begin{subfigure}{\textwidth}
        \centering
        \includegraphics[width=\textwidth]{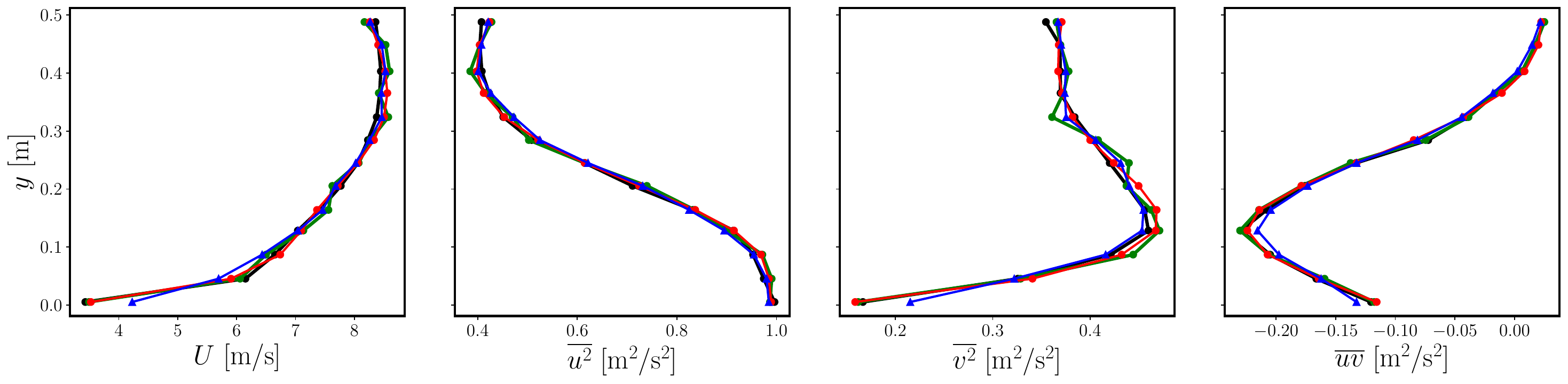}
    \end{subfigure}
    \begin{subfigure}{\textwidth}
        \centering
        \includegraphics[width=\textwidth]{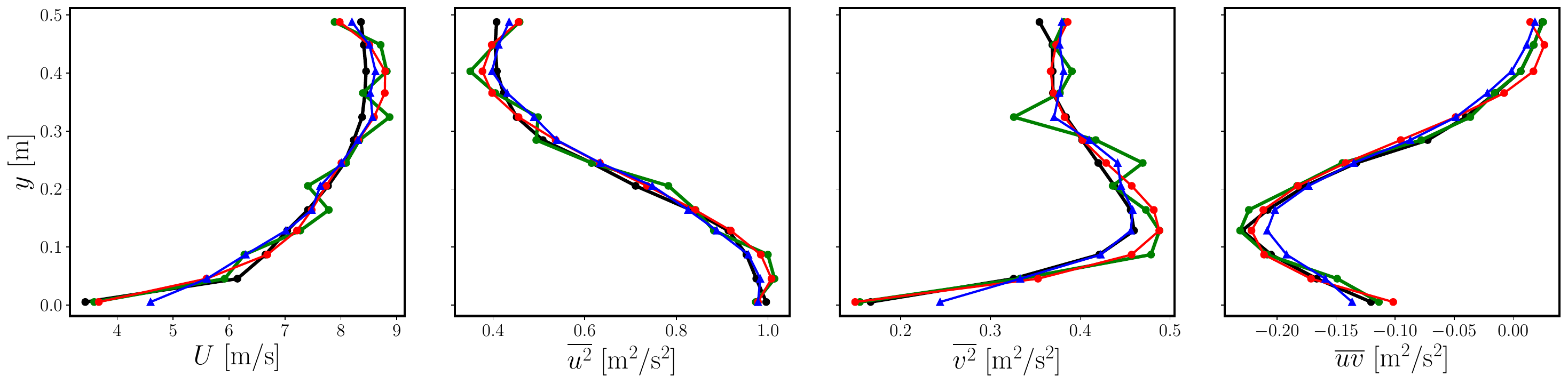}
    \end{subfigure}
    \begin{subfigure}{\textwidth}
        \centering
        \includegraphics[width=\textwidth]{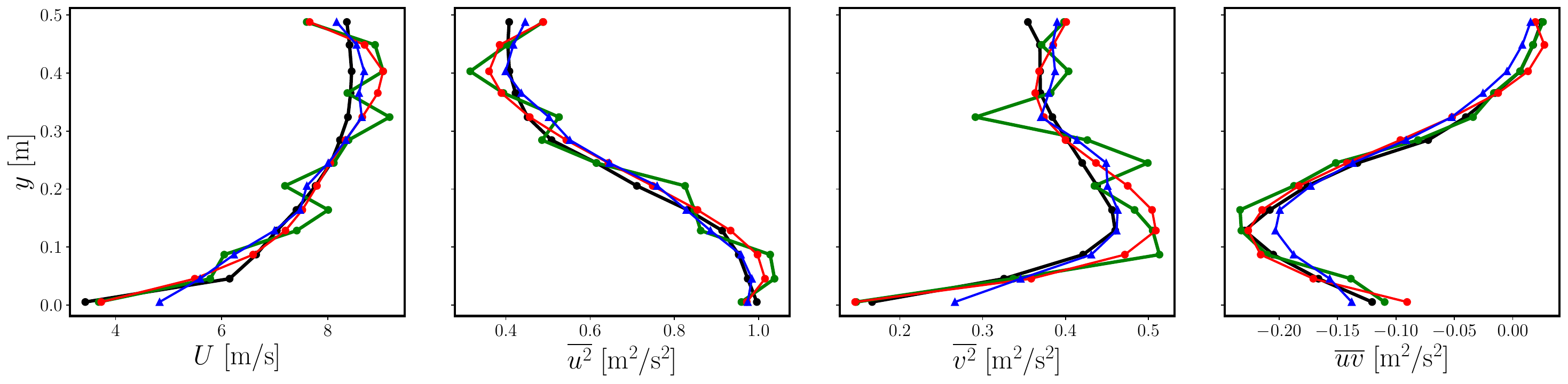}
    \end{subfigure}    
    \caption{De-noising results on (from left to right) mean flow, streamwise fluctuations, wall-normal fluctuations and Reynolds shear stress. (Black) original profile, (green) profile with superimposed synthetic noise, (red) profile after reducing the noise via PINNs and (blue) profile after reducing noise via least-square regression. From the top to the bottom we show results for the noise level of 2$\%$, 5$\%$ and 8$\%$, respectively.}
    \label{fig:denoising}
\end{figure}

In this section, we also begin with clean hot-wire measurements and then introduce Gaussian noise to simulate potential dispersion errors. We utilize the 3$\sigma$ criterion to adjust the standard deviation $\sigma$ in the Gaussian distribution, controlling the added noise on the profiles in terms of noise level, as previously demonstrated in the literature~\cite{vinuesa_eif}. For each scaled quantity $\phi^+$ in the profile, $\sigma$ is calculated as:
\begin{equation}
    \sigma = \frac{\rm N [\%]}{100} \cdot \frac{\phi^+}{3},
\end{equation}
where $\rm N$ represents the noise level to be added to the profile. It is important to note that we use a friction velocity $u_{\tau}$ of 0.48 m/s as provided in the original database~\cite{bayoan_et_al} to scale the quantities $\phi$ as  {$\phi^+$}, following the logarithmic law of the wall. We consider three noise level cases: 2\%, 5\%, and 8\%. As illustrated by the green lines in ~\cref{fig:denoising}, it can be observed that the synthetic profile becomes noisier when the quantity exhibits higher values. We will employ the same architecture as that discussed in \S\ref{sec:super_resolution}, which was selected following the hyper-parameter study detailed in~\ref{sec:tune_hotwire}, where we first search for the optimal architectures and subsequently tune the respective weights of the supervised and unsupervised loss. It is worth noting that we adopt the same architecture used in the super-resolution task as it performs overall best on both tasks.

Additionally, we utilize weighted least-square regression~\cite{selesnick} as a conventional de-noising method for comparison purposes. Denoting $\mathbf{y} \in \mathbb{R}^n$ as the noisy input and $\mathbf{\hat{y}} \in \mathbb{R}^n$ as the approximation of the clean output, the problem can be formulated as:
\begin{equation}
    \min_{\mathbf{\hat{y}}} \| \mathbf{y} - \mathbf{\hat{y}} \|^2_2 + \lambda \| \mathbf{D} \mathbf{\hat{y}} \|^2_2,
    \label{eq:de-noise}
\end{equation}
where $\mathbf{D}\mathbf{\hat{y}}$ represents the second-order differentiation of $\mathbf{\hat{y}}$, and $\lambda$ is a hyper-parameter controlling the weight of $\mathbf{\hat{y}}$ for de-noising the noisy profile. This formula yields the expression of the de-noised profile $\mathbf{\hat{y}}$ using least-square methods, given by:
\begin{equation}
    \mathbf{\hat{y}} = (\mathbf{I}  - \lambda \mathbf{D}^T\mathbf{D})^{-1} \mathbf{y},
\end{equation}
where $\mathbf{I}$ is the identity matrix of the same size as $\mathbf{D}$. Note that we adopt $\lambda$ values of 0.5, 1, and 1.5 for noise levels of 2\%, 5\%, and 8\%, respectively.

Table~\ref{tab:de-noise} summarizes the relative Euclidean norm of error obtained with respect to the clean profiles for the proposed noise levels. PINNs yield an average error $\overline{\epsilon}$ of 2.17\%, 5.07\%, and 6.99\% for noise levels of 2\%, 5\%, and 8\%, respectively. Our results indicate that in all cases, the noise is reduced, leading to a better agreement with the clean original data compared to the weighted least-square regression, which yields average errors of 3.52\%, 5.57\%, and 7.17\% for noise levels of 2\%, 5\%, and 8\%, respectively.
\begin{table}[ht]
    \centering
    \caption{Relative Euclidean norm of errors with respect to clean profile obtained at noise level (N) for 2$\%$, 5$\%$ and 8$\%$. The boldface denotes the lowest averaged error obtained in each case.}
        \begin{tabular}{ccccccc}
            \hline
            \hline
            & N (\%) &  $\epsilon_{U}$ (\%) & $\epsilon_{\overline{u^2}}$ (\%) & $\epsilon_{\overline{v^2}}$ (\%) & $\epsilon_{\overline{uv}}$ (\%) & $\overline{\epsilon}$ (\%) \\\hline
            \rule{0pt}{3ex}Noisy profile        & 2    &   1.63   & 2.19        & {3.13}  & {2.39}  & 2.34 \\  
            PINNs        & 2    &  {1.39}   & {1.40}        & {2.33} & 3.56 & {2.17} \\ 
            Least-square & 2    &   3.52    & 1.77         & 3.85   & 4.92 & 3.52 \\ \hline
            \rule{0pt}{3ex}Noisy profile        & 5    &   4.09   & 5.46        & {7.83}  &  {5.97}  & 5.84 \\
            PINNs        & 5    &  {3.39}   & 3.59        & {5.19} & 8.12 & {5.07} \\ 
            Least-square & 5    &   5.08    &{3.09}         & 6.29   & 7.81 & 5.57 \\ \hline
            \rule{0pt}{3ex}Noisy profile & 8    &   6.55            & 8.75        & 12.62  & 9.55  & 9.34 \\
            PINNs        & 8    &  {4.96}   & 5.46        & {8.16} & {9.40} & {6.99} \\ 
            Least-square & 8    &   6.01           & {4.14}         & 8.41   & 10.10 & 7.17 \\ 
            \hline
            \hline
        \end{tabular}       
    \label{tab:de-noise}
\end{table}

\subsubsection{Discussion of de-noising results}

As depicted in \cref{fig:denoising}, PINNs reduce the relative error with respect to the clean profile in all  {the} cases. Particularly, for the $N = 2\%$ case, the Gaussian noise added synthetically is $2.34\%$, and the PINNs correction leads to an average relative error of $2.17\%$. As the noise level increases to $5\%$, the initial synthetic error of $5.84\%$ is reduced to $5.07\%$. When the noise level rises to $8\%$, the initial synthetic error is reduced from $9.34\%$ to $6.99\%$, achieving a more significant noise reduction on the profile. It's noteworthy that the largest errors are observed close to the boundary-layer edge. This might be attributed to the smoother nature of the PINNs-based profile compared to the irregular noisy profile. Given the trend induced by the noisy points, the resulting curve leads to slightly higher errors closer to the freestream. Comparing with the results obtained by weighted least-square regression, it's evident that the profiles obtained by regression exhibit better agreement in the region closer to the freestream. However, it's important to note that the streamwise mean velocity and wall-normal fluctuation profiles denoised by weighted least-square regression demonstrate a distinct shift at the wall in all proposed cases, damaging the near-wall information. This observation implies that our PINNs method exhibits a more robust performance at the near-wall region than the linear method, which is particularly useful for turbulent boundary layer measurement.

\section{Summary and Conclusions}
\label{sec:conclusions}

Applications of physics-informed deep learning have been demonstrated for data augmentation in experimental fluid mechanics, where the measurement data are generally imperfect, noisy and have low resolution. We employ physics-informed neural networks (PINNs) for super-resolution of flow-field data both in time and space from a limited set of noisy measurements without having any high-resolution targets. Our objective is to solve the governing partial differential equations for an ill-posed problem, \ie unknown initial and boundary conditions, only having a few low-resolution and noisy measurements. To this end, we use a neural network to obtain a continuous solution of the problem, providing a physically-consistent prediction at any point in the solution domain. In particular, we learn a functional mapping parameterized by the neural network from spatial ($x$) and temporal ($t$) coordinates to the flow parameters, \ie velocity components and the pressure. We first demonstrate the applicability of PINNs to the super-resolution tasks in time using the canonical case of Burgers’ equation. We only take the data at the first and the last time-steps of the solution domain for supervised learning. This can be seen as an extensive super-resolution task in time using only two snapshots of the data for training. Our results show that the PINN model can accurately provide a continuous solution of the shock formation process with a relative Euclidean norm of errors of $\epsilon_u = 0.13\%$ . These results show an adequate performance of PINNs in such tasks and motivate their application to this type of problem in fluid mechanics. Next, we perform the super-resolution in time and space for the two-dimensional vortex shedding behind a circular cylinder at $Re_D = 100$ to highlight the ability of PINNs in flow-field super-resolution tasks from very low resolution and noisy measurements. We consider a time interval of almost one vortex-shedding cycle and only utilize three snapshots of low-resolution and noisy data for supervised learning. Our results indicate that a very good reconstruction of the temporal dynamics can be obtained even for the high-frequency modes, showing good performance of the PINN model in increasing the temporal resolution of the data and correcting the effect of experimental noise. We also apply PINNs for the reconstruction of a continuous solution for a minimal turbulent channel at $Re_{\mathrm{cl}} = 5,000$ from a limited set of measurements with different resolutions in time and space and assess the robustness of the model against noise. These results indicate that adequate predictions can be obtained from the PINN model by using only five low-resolution snapshots of the flow for supervised learning, leading to $\epsilon_u$ = 0.53\%, $\epsilon_v$ = 12.54\% and $\epsilon_w$ = 9.89\%. Very good results can also be achieved using only three snapshots, which leads to $\epsilon_u$ = 0.81\%, $\epsilon_v$ = 17.57\% and $\epsilon_w$ = 20.75\%. Based on the two simulation-based cases, a numerical trend that can be extrapolated to future application scenarios: PINNs exhibit better performance in resolving large-scale quantities compared to small-scale ones and in  {representing} streamwise quantities. This trend aligns with observations from experimental applications in the present study. We also show that the PINN model is robust against noise. Finally, we show that the PINNs framework is capable of improving the resolution and reducing the noise in an experimental dataset consisting of hot-wire anemometry measurements of a turbulent boundary layer. The present study demonstrates the potential of PINNs in the context of physics-informed data augmentation for experiments in fluid mechanics.

\section*{Acknowledgments}

RV acknowledges the financial support by the G\"oran Gustafsson foundation and ERC grant no. `2021-CoG-101043998, DEEPCONTROL'. Part of the analysis was carried out using computational resources provided by the  National Academic Infrastructure for Supercomputing in Sweden (NAISS).

\appendix
\section*{Appendix}
\setcounter{section}{1}
\subsection{Fine-tuning for two-dimensional vortex shedding behind a circular cylinder}
\label{sec:tune_vortex}
In this section, we report the hyper-parameter study conducted to obtain the architecture employed across \S\ref{sec:cylinder}. Note that these results correspond to the case with a noise level of $5\%$. We will consider variations of the number of hidden layers $l$, neurons per layer $n$, coefficients for the supervised ($\alpha$) and unsupervised ($\beta$) losses in equation~(\ref{eq:loss}), relative errors of all the statistics $\epsilon$, average relative error $\overline{\epsilon}$ and computational time required to train the model $t_c$. Note that $t_c$ is evaluated on a single central processing unit (CPU) from a machine with the following characteristics: AMD Ryzen 9 7950X with 16 cores, 32 threads, 4.5 Ghz and 62 GB RAM. In Table~\ref{tab:ap1_vortex} we show the first hyper-parameter study, where we change the values of $l$ and $n$ keeping $\alpha$ and $\beta$ constant. Based on this first study, we find that $l=4$ and $n=20$ yield the lowest average relative error, and proceed with the second study shown in Table~\ref{tab:ap2_vortex}, where we keep the previous values of $l$ and $n$ and systematically change the coefficients $\alpha$ and $\beta$. The final configuration has $\alpha=1$, $\beta=10$, and will be employed for this flow case.

\begin{table}[ht]
    \centering
    \caption{Hyper-parameter study for two-dimensional vortex shedding behind a circular cylinder fixing $\alpha=1$ and $\beta=10$. Boldface denotes the architecture with the lowest average relative error.}
    \vspace{9pt}
    \begin{tabular}{ccccccccc}
        \hline 
        \hline
        \vspace{5pt}
        
        $l$ & $n$ & $\alpha$& $\beta$ & $\epsilon_{u}$ [$\%$] & $\epsilon_{v}$ [$\%$] & $\epsilon_{p}$ [$\%$] & $t_c$ [s] & $\overline{\epsilon}$ [$\%$] \\
        \hline
4 & 20 & 1 & 10 & 3.96 & 8.66 & 19.06 & 414.06 & {10.56} \\
4 & 60 & 1 & 10 & 6.29 & 11.77 & 61.76 & 595.10 & 26.61  \\
4 & 100 & 1 & 10 & 7.25 & 14.16 & 97.69 & 1082.40 & 39.70  \\
6 & 20 & 1 & 10 &4.47 & 8.54 & 27.07 & 311.43 & 13.36\\ 
6 & 60 & 1 & 10 & 6.04 & 11.54 & 48.26 & 1070.28 & 21.95  \\
6 & 100 & 1 & 10 & 7.45 & 15.57 & 55.34 & 1948.66 & 26.12 \\
10 & 20 & 1 & 10 & 13.28 & 54.81 & 56.34 & 35.82 & 41.48  \\
10 & 60 & 1 & 10 & 5.76 & 15.74 & 23.44 & 31.19 & 14.98  \\
10 & 100 & 1 & 10 &8.77 & 16.44 & 48.44 & 4261.96 & 24.55 \\
         
    \hline
    \hline
    \end{tabular}
    \label{tab:ap1_vortex}
\end{table}

\begin{table}[ht]
    \centering
    \caption{Hyper-parameter study for two-dimensional vortex shedding behind a circular fixing $l=4$ and $n=20$. Boldface denotes the architecture with the lowest average relative error.}
    \vspace{9pt}
    \begin{tabular}{ccccccccc}
        \hline 
        \hline
        \vspace{5pt}
        
        $l$ & $n$ & $\alpha$& $\beta$ & $\epsilon_{u}$ [$\%$] & $\epsilon_{v}$ [$\%$] & $\epsilon_{p}$ [$\%$] & $t_c$ [s] & $\overline{\epsilon}$ [$\%$] \\
        \hline
4 & 20 & 1 & 1 & 4.27 & 11.09 & 21.62 & 167.3 & 12.33\\
4 & 20 & 1 & 5 & 4.35 & 8.52 & 21.7 & 151.74 & 11.52  \\
4 & 20 & 1 & 10 & 3.96 & 8.66 & 19.06 & 414.06 & {10.56}\\
4 & 20 & 5 & 1 & 5.94 & 12.36 & 21.67 & 224.39 & 13.32\\ 
4 & 20 & 5 & 5 & 4.59 & 9.67 & 24.94 & 131.02 & 13.07  \\
4 & 20 & 5 & 10 & 4.27 & 9.58 & 20.55 & 152.62 & 11.47 \\
4 & 20 & 10& 1 & 5.54 & 13.48 & 18.93 & 164.99 & 12.65  \\
4 & 20 & 10& 5 & 4.90 & 10.96 & 18.62 & 158.7 & 11.49 \\
4 & 20 & 10& 10 &4.52 & 8.81 & 23.45 & 101.97 & 12.26 \\
         
    \hline
    \hline
    \end{tabular}
    \label{tab:ap2_vortex}
\end{table}

\subsection{Fine-tuning for minimal turbulent channel}
\label{sec:tune_channel}
In this section, we report the hyper-parameter study conducted to obtain the architecture employed across \S\ref{sec:channel}. Note that these results correspond to case PINN--t5--s8. We conducted our hyper-parameter studies following the same procedure and utilizing the same machine for training as in the previous section. In Table~\ref{tab:ap1_channel} we show the first hyper-parameter study, where we change the values of $l$ and $n$ keeping $\alpha$ and $\beta$ constant. Based on this first study, we find that $l=10$ and $n=100$ yield the lowest average relative error, and proceed with the second study shown in Table~\ref{tab:ap2_channel}, where we keep the previous values of $l$ and $n$ and systematically change the coefficients $\alpha$ and $\beta$. The final configuration has $\alpha=1$, $\beta=1$, and will be employed for the minimal turbulent channel.

\begin{table}[ht]
    \centering
    \caption{Hyper-parameter study for turbulent minimal channel PINN--t5--s8 fixing $\alpha=1$ and $\beta=1$. Boldface denotes the architecture with the lowest average relative error.}
    \vspace{9pt}
    \begin{tabular}{ccccccccc}
        \hline 
        \hline
        \vspace{5pt}
        
        $l$ & $n$ & $\alpha$& $\beta$ & $\epsilon_{u}$ [$\%$] & $\epsilon_{v}$ [$\%$] & $\epsilon_{w}$ [$\%$] & $t_c$ [s] & $\overline{\epsilon}$ [$\%$] \\
        \hline
        4 & 20 & 1 & 1 & 3.14 & 64.45 & 53.16 & 231.48 & 40.25\\
        4 & 60 & 1 & 1 & 2.32 & 32.22 & 45.02 & 1828.71 & 26.52  \\
        4 & 100 & 1 & 1 & 2.05 & 30.41 & 45.07 & 2941.71 & 25.84\\
        6 & 20 & 1 & 1 & 2.35 & 42.86 & 47.83 & 771.99 & 31.01\\ 
        6 & 60 & 1 & 1 & 1.78 & 28.3 & 43.07 & 2690.33 & 24.38  \\
        6 & 100 & 1 & 1 & 1.65 & 26.46 & 42.29 & 5267.81 & 23.47 \\
        10 & 20 & 1 & 1 & 1.97 & 33.0 & 44.34 & 954.06 & 26.43 \\
        10 & 60 & 1 & 1 & 1.56 & 26.21 & 44.6 & 4938.71 & 24.12  \\
        10 & 100 & 1 & 1 & 1.87 & 22.86 & 44.15 & 4817.92 & {22.95}\\
        \hline
        \hline
        \end{tabular}
        \label{tab:ap1_channel}
\end{table}

\begin{table}[ht]
    \centering
    \caption{Hyper-parameter study for turbulent minimal channel PINN--t5--s8 fixing $l$ = 10 and $n$ = 100. Boldface denotes the architecture with the lowest average relative error.}
    \vspace{9pt}
    \begin{tabular}{ccccccccc}
        \hline 
        \hline
        
        \vspace{5pt}
        
        $l$ & $n$ & $\alpha$& $\beta$ & $\epsilon_{u}$ [$\%$] & $\epsilon_{v}$ [$\%$] & $\epsilon_{w}$ [$\%$] & $t_c$ [s] & $\overline{\epsilon}$ [$\%$] \\
        \hline
        100 & 10 & 1 & 1 & 1.87 & 22.86 & 44.15 & 4817.92 & {22.95}\\
        100&10&1&5&1.56&27.82&41.76&5196.15&23.71 \\
        100&10&1&10&1.68&28.29&48.7&5073.0&26.22 \\
        100&10&5&1&1.58&26.59&41.18&5781.06&23.12 \\
        100 & 10 & 5 & 5 & 1.69 & 27.16 & 45.44 & 6480.06 & 24.76 \\
        100&10&5&10&1.74&28.98&44.16&5347.4&24.96 \\
        100&10&10&1&1.67&27.22&46.18&5187.21&25.02 \\
        100&10&10&5&1.8&27.53&40.86&6684.72&23.4 \\
        100&10&10&10&1.78&27.42&44.14&5803.82&24.45 \\

    \hline
    \hline
    \end{tabular}
    \label{tab:ap2_channel}
\end{table}

\subsection{Fine-tuning for experimental application}
\label{sec:tune_hotwire}
In this section, we will report the hyper-parameter study conducted to obtain the architecture employed across \S\ref{sec:exp}. Note that these results correspond to the study of super-resolution, but the same architecture is used for de-noising. We conducted our hyper-parameter studies following the same procedure and utilizing the same machine for training as in the previous section. In Table~\ref{tab:ap1} we show the first hyper-parameter study, where we change the values of $l$ and $n$ keeping $\alpha$ and $\beta$ constant. Based on this first study, we find that $l=4$ and $n=40$ yield the lowest average relative error, and proceed with the second study shown in Table~\ref{tab:ap2}, where we keep the previous values of $l$ and $n$ and systematically change the coefficients $\alpha$ and $\beta$. The final configuration has $\alpha=10$, $\beta=1$, and will be employed for both super-resolution and de-noising tasks. 

\begin{table}[ht]
    \centering
    \caption{Hyper-parameter study for application to experimental dataset fixing $\alpha=10$ and $\beta=1$. Boldface denotes the architecture with the lowest average relative error.}
    \vspace{9pt}
    \begin{tabular}{cccccccccc}
        \hline 
        \hline
        \vspace{5pt}
        
        $l$ & $n$ & $\alpha$& $\beta$ & $\epsilon_{U}$ [$\%$] & $\epsilon_{\overline{u^2}}$ [$\%$] & $\epsilon_{\overline{v^2}}$ [$\%$] & $\epsilon_{\overline{uv}}$ [$\%$] & $t_c$ [s] & $\overline{\epsilon}$ [$\%$] \\
        \hline
2 & 40 & 10 & 1 & 1.57 & 0.76 & 1.23 & 3.69 & 6.17 & 1.81 \\
2 & 60 & 10 & 1 & 1.64 & 0.97 & 2.19 & 5.45 & 6.12 & 2.56 \\
2 & 80 & 10 & 1 & 1.49 & 0.76 & 1.94 & 3.01 & 6.31 & 1.80 \\
2 & 100 & 10 & 1 & 2.11 & 0.78 & 0.94 & 3.56 & 8.42 & 1.85 \\ 
4 & 40 & 10 & 1 & 1.46 & 0.84 & 1.16 & 2.97 & 7.41 & {\bf 1.61} \\
4 & 60 & 10 & 1 & 1.14 & 1.68 & 1.46 & 5.16 & 9.79 & 2.36 \\
4 & 80 & 10 & 1 & 1.49 & 0.86 & 1.81 & 3.69 & 9.40 & 1.96 \\
4 & 100 & 10 & 1 & 1.70 & 0.85 & 2.28 & 3.28 & 11.99 & 2.03 \\
6 & 40 & 10 & 1 & 1.19 & 0.84 & 1.24 & 4.07 & 10.38 & 1.83 \\
6 & 60 & 10 & 1 & 1.23 & 1.13 & 1.23 & 3.43 & 13.71 & 1.76 \\
6 & 80 & 10 & 1 & 1.50 & 0.88 & 0.93 & 3.28 & 15.49 & 1.65 \\
6 & 100 & 10 & 1 & 1.61 & 0.93 & 0.84 & 3.38 & 18.62 & 1.69 \\
8 & 40 & 10 & 1 & 1.53 & 1.03 & 1.27 & 3.71 & 14.47 & 1.88 \\
8 & 60 & 10 & 1 & 1.50 & 0.96 & 1.01 & 5.35 & 18.28 & 2.21 \\
8 & 80 & 10 & 1 & 1.80 & 0.95 & 2.34 & 3.51 & 20.84 & 2.15 \\
8  & 100 & 10 & 1 & 1.96 & 1.43 & 0.90 & 7.67 & 23.8 & 2.99 \\          
    \hline
    \hline
    \end{tabular}
    \label{tab:ap1}
\end{table}

\begin{table}[ht]
    \centering
    \caption{Hyper-parameter study for application to experimental dataset fixing $l=4$ and $n=40$. Boldface denotes the architecture with the lowest average relative error.}
    \vspace{9pt}
    \begin{tabular}{cccccccccc}
        \hline 
        \hline
        \vspace{5pt}
        
        $l$ & $n$ & $\alpha$& $\beta$ & $\epsilon_{U}$ [$\%$] & $\epsilon_{\overline{u^2}}$ [$\%$] & $\epsilon_{\overline{v^2}}$ [$\%$] & $\epsilon_{\overline{uv}}$ [$\%$] & $t_c$ [s] & $\overline{\epsilon}$ [$\%$] \\
        \hline
            4 & 40 & 1 & 1 & 1.36 & 1.00 & 1.65 & 3.78 & 9.01 & 1.95 \\
            4 & 40 & 1 & 2 & 1.13 & 1.13 & 1.01 & 3.92 & 8.50 & 1.80 \\
            4 & 40 & 1 & 5 & 1.24 & 0.84 & 0.89 & 3.61 & 9.71 & 1.64 \\
            4 & 40 & 1 & 10 & 1.60&  0.87 & 1.24 & 4.17 & 8.43 & 1.97 \\
            4 & 40 & 1 & 20 & 1.69 & 0.97 & 1.16 & 3.58 & 8.43 & 1.85 \\
            4 & 40 & 2 & 1 & 1.53 & 1.00 & 1.51 & 3.85 & 7.32 & 1.97 \\
            4 & 40 & 2 & 2 & 1.23 & 0.85 & 1.36 & 3.59 & 8.11 & 1.76 \\
            4 & 40 & 2 & 5 & 1.59 &  0.93 & 0.91 & 3.64 & 8.54 & 1.77 \\
            4 & 40 & 2 & 10 & 1.04 & 0.79 & 0.79 & 3.62 & 10.29 & 1.56 \\
            4 & 40 & 2 & 20 & 1.43 & 0.86 & 0.91 & 4.08 & 8.61 & 1.82 \\
            4 & 40 & 5 & 1 & 1.47 & 0.90 & 1.11 & 4.14 & 7.94 & 1.91 \\
            4 & 40 & 5 & 2 & 1.51 & 0.98 & 1.28 & 3.27 & 9.11 & 1.76 \\
            4 & 40 & 5 & 5 & 1.46 & 0.77 & 0.95 & 3.46 & 7.47 & 1.66 \\
            4 & 40 & 5 & 10 & 1.38 & 1.06 & 1.46 & 3.45 & 8.28 & 1.84 \\
            4 & 40 & 5 & 20 & 1.22 & 0.83 & 1.40 & 3.21 & 9.13 & 1.67 \\
            4 & 40 & 10 & 1 & 1.46 & 0.84 & 1.16 & 2.97 & 7.41 & {\bf 1.61} \\
            4 & 40 & 10 & 2 & 1.71 & 0.91 & 0.86 & 3.41 & 7.74 & 1.72 \\
            4 & 40 & 10 & 5 & 1.66 & 0.84 & 1.17 & 3.01 & 6.92 & 1.67 \\     
    \hline
    \hline
    \end{tabular}
    \label{tab:ap2}
\end{table}

\subsection{Discussion on hyper-parameter selection and generalizability for future applications}
\begin{figure}[ht]
    \centering
    \begin{subfigure}{0.45\linewidth}
        \centering
        \includegraphics[width=\textwidth]{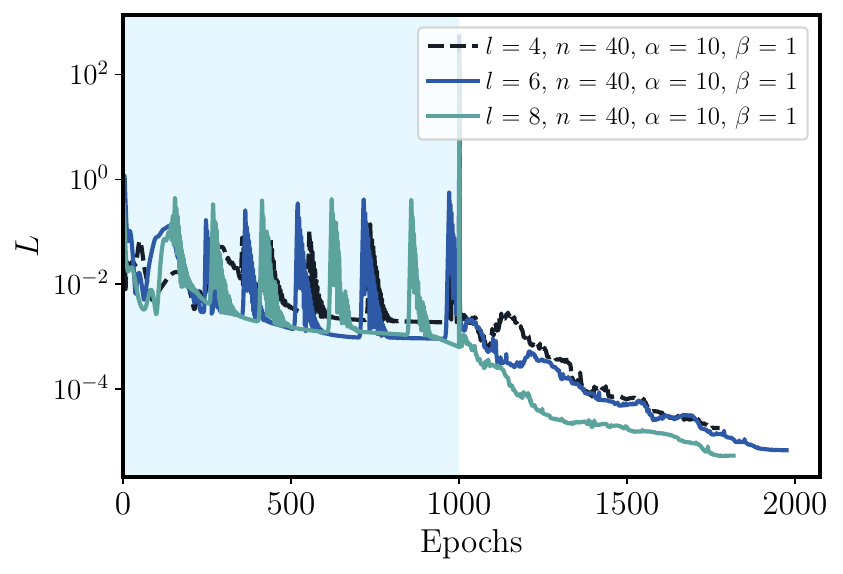}
    \end{subfigure}
    \quad
    \begin{subfigure}{0.45\linewidth}
        \centering
        \includegraphics[width=\textwidth]{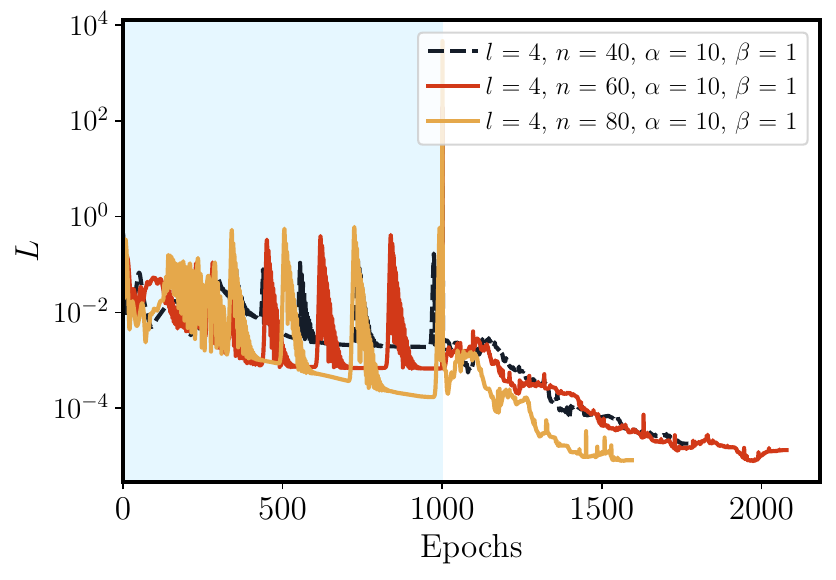}
    \end{subfigure}
    \quad   
    \begin{subfigure}{0.45\linewidth}
        \centering
        \includegraphics[width=\textwidth]{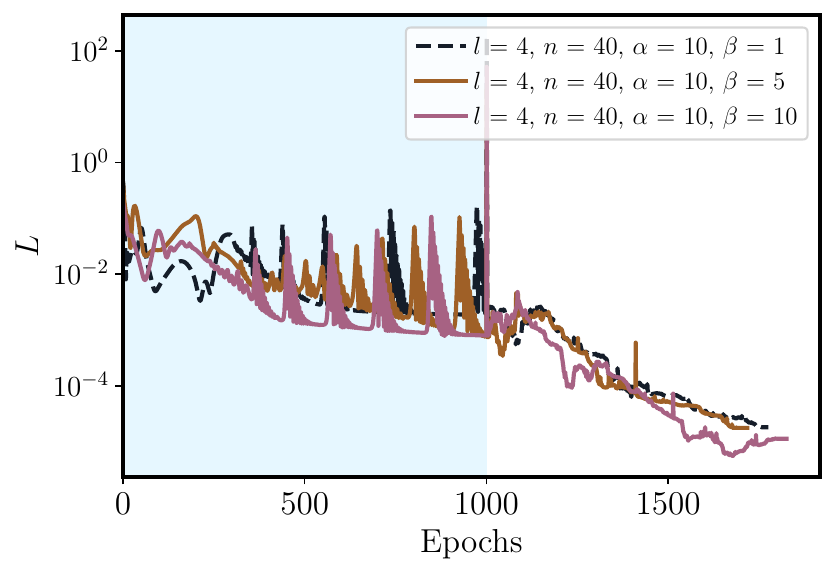}
    \end{subfigure}
    \quad
    \begin{subfigure}{0.45\linewidth}
        \centering
        \includegraphics[width=\textwidth]{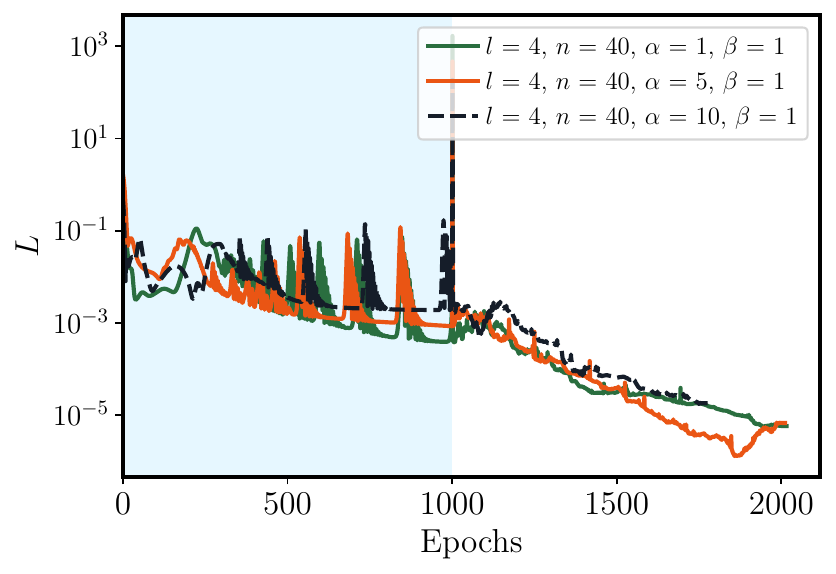}
    \end{subfigure}
    \caption{The convergence of the total loss ($L$) defined by~\cref{eq:loss} for hyper-parameter studies  in experimental application. We demonstrate the loss evolution of models fixing number of layers $l$ (upper left), number of neurons $n$ (upper right), weight of supervised-learning loss $\alpha$ (lower left) and weight of unsupervised-learning loss $\beta$ (lower right). Note that the black dashed line denotes the loss evolution obtained by the adopted model and the blue region denotes the epochs using the Adam optimizer.}
    \label{fig:param_hotwire}
\end{figure}

In this section, we discuss some guidelines for adopting the employed architecture and extending our conclusions for future applications. Since we considered several different tasks and flow cases, the employed architectures should always be fine-tuned according to the particular study to be conducted based on hyperparameter studies. One should begin by fine-tuning the architecture to achieve convergence of the loss and a lower evaluation error if possible. Subsequently, one can further fine-tune the weights of the supervised and unsupervised losses for the optimal architecture to enhance the model's performance. Note that defining the weights of the losses before fine-tuning is possible since they usually depend on the extent to which the results rely on the input data.
If there is no reference data available for comparison, which is a common scenario in many experimental applications, the convergence of the total loss $L$ might serve as a metric to evaluate the reliability of the results. \Cref{fig:param_hotwire} depicts the evolution of the loss when fixing different parameters. It can be observed that the total losses converge to similar magnitudes for different models. Given that a certain degree of deviation in the averaged relative error reported in \Cref{tab:ap1} and \Cref{tab:ap2} between two individual models is acceptable, the magnitude of the loss at the convergence stage can be considered a reliable metric for performance evaluation.

Furthermore, it is worth noting that the complexity of the adopted architecture is not necessarily proportional to the scale of the data but rather depends on the degrees of freedom of the unknowns. In the present study, the employed model for two-dimensional vortex shedding has three outputs ($u$, $v$ and $p$), while in experimental applications, there are six outputs ($u$, $v$, $uu$, $vv$, $uv$ and $p$). This is twice the number of outputs compared to the previous one, although the two-dimensional flow case is more complex. Therefore, even though both models comprise four layers, the architecture for experimental applications adopts 40 neurons ($n$) for each layer, while 20 neurons are used for the vortex-shedding case in each layer.

\FloatBarrier
\section*{References}

\bibliographystyle{jphysicsB}
\bibliography{References}

\end{document}